\documentclass[a4paper,11pt]{article}
\usepackage{amsmath}
\usepackage{amssymb}
\usepackage{color}
\usepackage{cite}
\usepackage{hyperref}
\usepackage{graphicx}
 \usepackage{leftidx}
 \usepackage{subfig}

\makeatletter
\@addtoreset{equation}{section}
\renewcommand{\theequation}{\thesection.\@arabic\c@equation}
\makeatother
\makeatletter
\renewcommand\appendix{\par
  \setcounter{section}{0}%
  \setcounter{subsection}{0}%
  \gdef\thesection{Appendix \@Alph\c@section }
  \renewcommand{\theequation}
  {\Alph{section}.\arabic{equation}}
}
\makeatother

\def \be {\begin{equation}}
\def \ee {\end{equation}}
\def \ba {\begin{array}}
\def \ea {\end{array}}
\def \bea{\begin{eqnarray}}
\def \eea{\end{eqnarray}}

\def \g {\gamma}

\def \p {\partial}

\def \nn {\nonumber}

\def \hs {\hspace}

\def \inf {\infty}

\def \Tr {{\textrm{Tr}}}

\def \cL{\cal L}
\def \cO{\cal O}

\title{\textbf{1-loop partition function in $AdS_3/CFT_2$}}
\author{
Bin Chen$^{1,2,3}$\footnote{bchen01@pku.edu.cn}\,
and
Jie-qiang Wu$^{1}$\footnote{jieqiangwu@pku.edu.cn}
}
\date{}

\begin{document}

\maketitle

\begin{center}
{\it
$^{1}$Department of Physics and State Key Laboratory of Nuclear Physics and Technology, Peking University, Beijing 100871, P.R.\! China\\
\vspace{2mm}
$^{2}$Collaborative Innovation Center of Quantum Matter, 5 Yiheyuan Rd, \\Beijing 100871, P.~R.~China\\
$^{3}$Center for High Energy Physics, Peking University, 5 Yiheyuan Rd, \\Beijing 100871, P.~R.~China
}
\vspace{10mm}
\end{center}

\begin{abstract}

The 1-loop partition function of the handlebody solutions in the AdS$_3$ gravity have been derived some years ago using the heat kernel techniques and the method of images. In the semiclassical limit, such partition function should correspond to the order $O (c^0)$ part in the partition function of dual conformal field theory(CFT) on the boundary Riemann surface.  The higher genus partition function could be computed by the multi-point functions in the Riemann sphere via sewing prescription.  In the large central charge limit, the CFT is effectively free in the sense that to the leading order of $c$ the multi-point function is further simplified to be a  summation over the products of two-point functions of single-particle states. Correspondingly in the bulk, the graviton is freely propagating without interaction. Furthermore the product of the two-point functions may define the links, each of which is in one-to-one correspondence with the  conjugacy class of the Schottky group of the Riemann surface. Moreover, the value of a link is determined by the multiplier of the element in the conjugacy class. This allows us to reproduce exactly the gravitational 1-loop partition function. The proof can be generalized to the higher spin gravity and its dual CFT.

\end{abstract}

\baselineskip 18pt
\thispagestyle{empty}

\newpage

\section{Introduction}


The AdS$_3$  gravity, whose action includes a negative cosmological constant besides the Einstein-Hilbert term, provides a new angle to understand the AdS/CFT correspondence\cite{Maldacena:1997re}. It looks trivial as there is no local bulk degree of freedom, but it actually has global degrees of freedom. As shown in  \cite{Brown:1986nw} ,  under appropriate boundary conditions the asymptotic symmetry of the AdS$_3$ spacetime in the theory is generated by two copies of Virasoro algebra with the central charge
\be
c=\frac{3l}{2G}, \label{central}
\ee
where $l$ is the AdS radius and $G$ is the three-dimensional gravitational coupling constant. This suggests that there are boundary degrees of freedom which may describe the physics in the bulk. The AdS$_3$/CFT$_2$ correspondence  states that the quantum gravity in AdS$_3$ is dual to a two-dimensional(2D) CFT with the central charge (\ref{central}). In this correspondence, the BTZ black hole \cite{Banados:1992wn,Banados:1992gq} is dual to the highly excited states in CFT, and its macroscopic Bekenstein-Hawking entropy  could be counted by the degeneracy of the excited states and therefore has a microscopic interpretation\cite{Strominger:1997eq}. Moreover, the AdS$_3$ gravity could be topological in nature. It was found in \cite{Achucarro:1987vz} that its action in the first order formulation could be  written in forms of the Chern-Simons action. This raised the proposal that the AdS$_3$ gravity could be equivalent to a Chern-Simons theory with a gauge group $SL(2,C)$\cite{Witten:1988hc,Witten:2007kt}.  The Chern-Simons formulation of the AdS$_3$ gravity has been generalized to include the higher spin fields\cite{Blencowe:1988gj,Bergshoeff:1989ns} and led to the correspondence between the higher spin AdS$_3$ gravity and 2D CFT with ${\cal W}$ symmetry\cite{Campoleoni:2010zq,Henneaux:2010xg}.

Even though a precise definition of AdS$_3$ quantum gravity has not been well-established, its semiclassical limit is well-understood. In the AdS$_3$ gravity, all classical solutions are locally maximal symmetric and could be obtained as the quotients of the global AdS$_3$ by the subgroup of the isometry group $SL(2,C)$ \cite{Banados:1992gq}. Consequently the path integral of the AdS$_3$ gravity could be well defined in principle. In the semiclassical regime, the partition function should include the contributions from  all the saddle points. Among all semi-classical solutions, the handlebody solutions have been best understood. This class of solutions could be obtained as the quotients  of the AdS$_3$ spacetime by the Schottky group. At the asymptotic boundary, the configuration is a Riemann surface, which could be uniformized by the Schottky group. It was shown in \cite{Krasnov:2000zq} that the regularized semi-classical action of the handlebody solution could be described by a Liouville type action, the so-called Zograf-Takhtajan(ZT) action\cite{Zagraf:1988}\footnote{For the study on the semiclassical action of other hyperbolic solutions see\cite{Takhtajan:2002cc,Yin:2007at}.}. Furthermore, the 1-loop partition function of the handlebody solutions has been conjectured to be\cite{Yin:2007gv}
 \be\label{1loop} \log Z|_{\mbox{1-loop}}=-\sum_{\gamma}\sum_{m=2}^{\inf}\log|1-q_{\gamma}^m|, \ee
where $\gamma$ is the primitive conjugacy class of the Schottky group and $q_{\gamma}^{-\frac{1}{2}}$ is the larger eigenvalue of $ \gamma$. This relation has been obtained by the direct computation in the gravity by using the heat kernel techniques and the method of images\cite{Giombi:2008vd}.

On the dual CFT side, the semiclassical gravity corresponds to the large central charge limit. Despite of many efforts (see for example \cite{Witten:2007kt}), the explicit construction of the CFT is not clear yet.  It is expected to have a sparse light spectrum in the large $c$ limit\cite{Hartman:2013mia}\cite{Hartman:2014oaa}. To the leading order of $c$ in  the partition function,  the light spectrum dominate the contribution, while the heavy operators only contribute nonperturbatively  as $e^{-c}$. The vacuum module plays a special role. It is universal for all CFT, includes the stress tensor and its descendants. In the AdS$_3$/CFT$_2$ correspondence, the stress tensor is dual to the massless graviton in the bulk. In fact, it was shown in \cite{Maloney:2007ud} that  the genus-$1$ partition function is 1-loop exact and the contribution comes purely from the vacuum module. 

The recent study on the R\'enyi and entanglement entropy revives the AdS$_3$/CFT$_2$ correspondence.  The multi-interval  R\'enyi entropy of a 2D CFT is determined by the partition function on a higher genus Riemann surface resulted from the replica trick. For the CFT dual to the AdS$_3$ gravity, the CFT partition function should be equal to the partition function of corresponding gravitational configuration ending on the Riemann surface at the AdS boundary. In the large $c$ limit, the leading contribution of the R\'enyi entropy, which is dominated by the vacuum module\cite{Hartman:2013mia},  is equal to  the ZT action\cite{Faulkner:2013yia}. This leads to the proof of the Ryu-Takayanagi formula for the holographic entanglement entropy\cite{Ryu:2006bv,Ryu:2006ef}. More interestingly, from the study on the R\'enyi entropy of double intervals with a small cross-ratio and the single interval on a torus\cite{Headrick:2010zt,Barrella:2013wja,Chen:2013kpa,Chen:2013dxa,Perlmutter:2013paa,Chen:2014kja,Beccaria:2014lqa,Chen:2014unl,Chen:2015kua,Chen:2015uia}, it turns out that the holographic computation is even correct at 1-loop level. Now the next-leading contribution to the R\'enyi entropy from the vacuum module is dual to the 1-loop graviton partition function (\ref{1loop}) of corresponding gravitational configuration.  More generically, one may find the following picture. For a handlebody instanton whose boundary is a higher genus Riemann surface, the partition function of the instanton including the quantum corrections should be exactly the same as the partition function of the Riemann surface in the dual CFT in the large $c$ limit.  In this paper, we try to prove that the 1-loop partition function (\ref{1loop}) for any handlebody solution agrees exactly with the next-leading $c$ part of the partition function in CFT.


To prove the 1-loop partition function (\ref{1loop}) for any handlebody solution, we need to compute the partition function of a higher genus Riemann surface in CFT. Here we use the gluing prescription to compute the higher genus partition function\cite{Segal:2002ei,Yin:2007gv,Gaberdiel:2010jf,Headrick:2015gba}.
Every compact Riemann surface could be described by the Schottky uniformization. For a genus $g$ Riemann surface, the Schottky uniformization allows us to identify $g$ pairs of   nonintersecting circles in the Riemann sphere. In CFT language, the identification  is equivalent to cut open a handle and insert a complete set of states there. On the Riemann sphere, this means that one has to insert pairs of the vertex operators at the fixed points in  the pairwise circles. As a result, the partition function of a  genus-$g$ Riemann surface is the summation of  $2g$-point functions on the Riemann sphere. As there is a uniformization map from the Riemann surface to the Riemann sphere, the resulting conformal anomaly is proportional to the central charge, therefore  the linear $c$ contribution in the partition function is captured purely by the ZT action. The sub-leading contribution is encoded in the $2g$-point functions
\be
Z_g\mid_z=\sum_{m_1,m_2,...m_g} \langle
 ~\leftidx{^{{\cL}_1}}{\bar{O}}{^{(1)}_{m_1}}O_{m_1}^{(1)} ~\leftidx{^{{\cL}_2}}{\bar{O}}{^{(2)}_{m_2}}O_{m_2}^{(2)}~~...
  ~\leftidx{^{{\cL}_g}}{\bar{O}}{^{(g)}_{m_g}} ~O_{m_g}^{(g)}\rangle, \ee
where  $m_1,m_2,...m_g$ denote the summation of all of the states on the circles $C_1,C_2,...C_g$ and $C'_1,C'_2,...C'_g$,  ${\cL}_i$ denotes the Schottky generator identifying $C_i$ and $C'_i$, and $O^{(i)}_{m_i}, \leftidx{^{{\cL}_i}}{\bar{O}}{^{(i)}_{m_i}}$ are the vertex operators corresponding to the same state but being inserted at the different fixed points of the generator ${\cL}_i$. As we are interested in the next-leading contribution  in the large $c$ limit, the computation of $2g$-point functions is very much simplified.  

One essential fact is that the CFT in the large $c$ limit becomes effectively free, which means that the multi-point function is dominated by the product of two-point functions of single-particle states. First of all, a general state in the vacuum module could be of the form
\be
\prod_{m=2}{\hat L}_{-m}^{r_m}|0>, \label{general}
\ee
where ${\hat L}_{-m}$'s are the normalized Virasoro generators, $r_m$'s  are non-negative integers. In the large $c$ limit, different states are orthogonal to each other, and all of the states constitute a complete set. 
Every state ${\hat L}_{-m}|0>$ is a one-particle state as it could be constructed as $(L_{-1})^{m-2}|T>$, where $|T>={\hat L}_{-2}|0>$. From the state-operator correspondence, the corresponding vertex operator of ${\hat L}_{-m}|0>$ is of a form  $V_m \sim \p^{m-2}T$ so that it duals to a single graviton in the bulk.  The particle number of a general state (\ref{general}) is $r=\sum r_m$. For a particle-$r$ state, the normalized vertex operator is just the normal ordered product of single-particle operators
\be
{\hat O}=:\prod_{j=1}^r V_{m_j}: .
\ee
Secondly, the leading contribution in the $2g$-point functions on the Riemann sphere is of order $O(c^0)$, so that a $2g$-point function is dominated by the products of the two-point functions between the single-particle operators. 
Moreover the products of two-point functions may define various links. By using $SL(2,C)$ transformations and the reduced completeness condition, the value of a link is captured by  the correlator of two single-particle vertex operators being related by an element in the Schottky group. Consequently the value of the link is only determined by the multiplier of the element. More interestingly, it turns out that every oriented link is actually in one-to-one correspondence with the conjugacy class of the Schottky group. This paves the way to prove the 1-loop partition function (\ref{1loop}) for any handlebody solution by taking into account all possible combination of the products of two-point functions in $Z_g$.

In the next section, after  briefly reviewing the Schottky uniformization, we discuss how to compute the partition function of a CFT on a higher genus Riemann surface.  In section 3, we prove the relation (\ref{1loop}) for any handlebody solution for pure AdS$_3$ gravity. We discuss the states in the vacuum module of the CFT in the large $c$ limit and the corresponding vertex operators. As a warm up, we reconsider the genus-1 partition function. Then we move to the higher genus partition function. In section 4,   we generalize our study to the CFT with ${\cal W}$ symmetry. We end with the conclusion and discussion.


\section{Schottky uniformization and the partition function}

In 3D AdS gravity, all the classical solutions could be obtained as the quotient of the global AdS$_3$ spacetime by a subgroup of the isometry group $SL(2,C)$.  In this work, we focus on the handlebody solutions whose asymptotic boundaries are compact Riemann surface. For the handlebody solutions, the subgroup is actually a Schottky group. In general, for a handlebody solution, the boundary Riemann surface is of higher genus.

From AdS/CFT correspondence, the partition function of AdS$_3$ quantum gravity should correspond to the partition function of higher genus Riemann surface in the dual CFT. In the large central charge limit, the semiclassical gravitational action is captured by the leading $c$ terms in the CFT partition function. In the large $c$ CFT, the leading contribution is determined by the conformal anomaly and the Schottky uniformization. For the next-leading contribution, it could be read from the summation of the multi-point functions on the Riemann sphere via the sewing prescription.

In this section, we first give a brief review on Schottky uniformization,  mainly basing on the work\cite{Krasnov:2000zq}. Then we discuss  how to compute a higher genus  partition function using the sewing prescription.

\subsection{Schottky uniformization}

Every compact Riemann surface can be described by a  Schottky uniformization. For a genus-$g$ Riemann surface ${\cal{M}}$, it can be represented by the quotient ${\cal{M}}=\Omega/\Gamma$, where $\Omega$ is the full complex plane plus the point of infinity with the fixed points of $\Gamma$ being removed, and $\Gamma$ is the Schottky group freely generated by $g$ loxodromic $SL(2,C)$ elements ${\cL}_i$. $\Omega$ is called the region of discontinuity of $\Gamma$. Moreover, it is convenient to introduce the fundamental region to describe the Schottky group. A fundamental region $D$ is a subset of $\Omega$, such that the interior points in $D$ are not $\Gamma$ equivalent to each other. One may choose $2g$ non-intersecting circles $C_1,C_2,...C_g$ and $C_1^{'},C_2^{'},...C_g^{'}$ in the Riemann sphere such that all circles lie to the exterior of each other. The loxodromic element ${\cL}_i$(${\cL}_i^{-1}$) maps $C_i$ to $C'_i$ such that the outer(inner) part of $C_i$ is mapped to the inner(outer) part of $C_i^{'}$. Then the fundamental region is the part of the Riemann sphere exterior to all the circles, and its quotient is a compact Riemann surface of genus $g$.  Each element ${\cL}_i$ is  an $SL(2,C)$ matrix , and it is represented by the action
\be
\frac{{\cL}_i(z)-a_i}{{\cL}_i(z)-r_i}=p_i\frac{z-a_i}{z-r_i}. \label{Li}
\ee
where $a_i$ and $r_i$ are respectively the attracting and repelling fixed points, $0<|p_i|<1$ is the multiplier. The eigenvalues of the matrix ${\cL}_i$ are $\sqrt{p_i}$ and $\sqrt{p_i^{-1}}$. Therefore each generator ${\cL}_i$ is completely characterized by the fixed points $a_i,r_i$ and the multiplier $p_i$. Among $3g$ complex parameters $a_i, b_i$ and $p_i, i=1,\cdots, g$, one can fix three of them by using Mobius transformation. The Schottky group satisfying the above conditions are called   a normalized and marked Schottky group.
The remaining $3g-3$ parameters parametrize the Schottky space of genus $g$.

One may define the map
\be
\g_{a_i,r_i}(z)=\frac{r_iz+a_i}{z+1} \label{gammai}
\ee
such that $\g_{a_i,r_i}(0)=a_i,\g_{a_i,r_i}(\infty)=r_i$. It maps the standard unit circle centered at the origin to the circle $C_i$. With $\g_p(z)\equiv pz$, a Schottky generator ${\cL}_i$ in (\ref{Li}) is just
\be
{\cL}_i=\g_{a_i,r_i}\g_{p_i}\g^{-1}_{a_i,r_i}.
\ee
 Actually every Schottky generator could be constructed in this way.


The uniformization map from the region of discontinuity to the Riemann surface could be determined by the help of a second order differential
equation
\be\label{diff} \psi^{''}(u)+\frac{1}{2}R^{S}(u)\psi(u)=0, \ee
where $R^{S}(u)$ is the projective connection on a marked Riemann surface and $\psi$ could be taken as a multi-valued differential on ${\cal{M}}$ of order $-1/2$. The ratio of two independent solutions $\psi_1$ and $\psi_2$ of the Fuchsian equation (\ref{diff})
\be\label{con} z=\frac{\psi_1(u)}{\psi_2(u)} \ee
gives the map. By imposing appropriate monodromy condition on the cycles of the fundamental group of the Riemann surface, one can find the generators of the Schottky group. For a general higher genus Riemann surface, this is a very difficult problem.  However, for the Riemann surface resulted  from the replica trick in computing the R\'enyi entropy, the problem has been solved explicitly in a perturbative way in the double-interval case\cite{Faulkner:2013yia,Barrella:2013wja} and single interval on a torus case \cite{Barrella:2013wja}\cite{Chen:2015kua}.


\subsection{Partition function on higher genus Riemann surface}

We would like to compute the partition function of a large $c$ CFT on a higher genus CFT.  It turns out that the leading $c$ contribution is captured by the Zograf-Takhtajan(ZT) action. On any compact Riemann surface of genus greater than $1$, there is a so-called Poincar\'e metric, which is a unique complete metric of constant negative curvature $-1$
\be
d\hat s^2=\frac{dt d\bar t}{({\mbox{Im}}( t))^2}.
\ee
Such a metric is related to the flat metric on the complex plane by a conformal transformation
\be
d\hat s^2=e^{2\phi_s(z,\bar z)}dz d\bar z
\ee
where $\phi_s$ is a real field on the Riemann sphere. The constant curvature condition requires that the field satisfy the Liouville equation
\be
\p_z\p_{\bar z}\phi_s=\frac{1}{2}e^{2\phi_s}.
\ee
This equation is the Euler-Lagrange equation of the ZT action defined on the fundamental region in the Schottky uniformization\cite{Zagraf:1988}
\be
S_{ZT}[\phi_s]=-\frac{c}{24\pi }\int\int_D\frac{i}{2}dz\wedge d\bar z\left(4\p_z\phi_s\p_{\bar z}\phi_s+\frac{1}{2}e^{2\phi_s}\right)+\mbox{boundary terms}.
\ee
This action is a Liouville action with boundary terms. The action evaluated on the solution of the Liouville equation gives exactly the regulated AdS$_3$ gravitational action of the
corresponding gravitational configuration. Actually this relation helps us to fix the overall factor in the above action. Moreover the ZT action captures the conformal anomaly and depends only on the choice of the metric in a fixed conformal class\cite{Krasnov:2000zq}.
Under the conformal transformation, the partition function on the Riemann surface is related to the one on the Riemann sphere via
\be\label{conformal} Z\mid_u=e^{-S_{ZT}} Z\mid_z.\ee
It is remarkable that the ZT action captures the whole leading contribution in the partition function in the large $c$ limit.



The partition function on a higher genus Riemann surface can be computed using gluing prescription, following Segal's approach to conformal field theory\cite{Segal:2002ei}. As nicely reviewed in the appendix C of \cite{Gaberdiel:2010jf}, the partition function is defined to be the summation of $2g$-point functions on the Riemann sphere
\be
Z_g=\sum_{\phi_i,\psi_i \in {\cal H}}\prod_{i=1}^{g}G^{-1}_{\phi_i\psi_i}\langle \prod_{i=1}^{g}\phi_i[C_i]\psi_i[C'_i]\rangle_D, \label{Zg}
\ee
where $D$ is the fundamental region with boundary $\p D=\cup_i(C_i\cup C'_i)$. Here $\phi_i, \psi_i$ are the states in the Hilbert space $\cal H$, and $\phi_i[C_i]$ denote the states associated with the boundary circle $C_i$. The circle $C_i$ could be related to the standard circle around the origin by a Mobius transformation (\ref{gammai}): $C_i=\g_{a_i,r_i} C$. To simplify the notation, we will write $\g_{a_i,r_i}=\g_i$. Due to the state-operator correspondence, the states on the circle $C_i$ is created by the vertex operator at $\g_i(0)$. More precisely, there is a correspondence
\be
\phi_i[C_i] \rightarrow V(U(\g_i)p_i^{L_0}\phi_i, a_i),\label{idenphi}
\ee
where the operator $U$ is
\be
U(\g)=\g'(0)^{L_0}e^{L_1\frac{\g''(0)}{2\g'(0)}}.
\ee
Moreover,  the state on the circle $C'_i$ corresponds to the vertex operator
\be\label{out}
\psi_i[C'_i] \rightarrow V(U(\g_i\hat \g)\psi_i, r_i)
\ee
where $\hat \g \equiv 1/z$ maps the origin to the infinity.
In (\ref{Zg}), $G_{\phi\psi}$ is the metric on the space of  the states
\be
G_{\phi\psi}=\lim_{z\to \infty}\langle V(z^{2L_0}e^{zL_1}\psi,z)V(\phi,0)\rangle.
\ee
With the vertex operators, the partition function (\ref{Zg}) is changed to the summation over $2g$-point functions of the vertex operators inserted at $2g$ fixed points
\be
Z_g=\sum_{\phi_i,\psi_i \in {\cal H}}\prod_{i=1}^{g}G^{-1}_{\phi_i\psi_i}\langle \prod_{i=1}^{g}V(U(\g_i)p_i^{L_0}\phi_i, a_i)V(U(\g_i\hat \g)\psi_i, r_i)\rangle, \label{ZgV}
\ee

The relation (\ref{Zg}) could be understood in the following way: one can insert a complete set of states in the Hilbert space at each pair of the circles $C_i$ and $C'_i$, which are related by the Schottky generator ${\cL}_i$, and compute all the possible $2g$-point functions of corresponding vertex operators on the Riemann sphere. One may apply this relation to compute the partition function of any CFT on a higher genus Riemann surface.
The computation could be simplified if one can choose a complete set of orthogonal state basis, in which case the metric on the space of the states becomes trivial
\be \langle \bar{O}_{m'}\mid O_{m} \rangle=\lim_{z\to \infty}\langle V(z^{2L_0}e^{zL_1}O_{m'},z)V(O_m,0)\rangle=\delta_{mm^{'}}. \ee

Let us reconsider the genus-1 partition function in a CFT.  In this case, the partition function is decomposed into two-point functions
\be
Z_1=\sum_{\phi,\psi\in {\cal H}}G^{-1}_{\phi\psi}\langle  V(U(\g_1)p_1^{L_0}\phi, a_1)V(U(\g_1\hat \g)\psi, r_1)\rangle.
\ee
As the two-point function is conformal invariant, we may apply a $SL(2,C)$ transformation $\g^{-1}_1$ to the two-point functions and get
\be
Z_1= \lim_{z\to \infty}\sum_{\phi,\psi\in {\cal H}}G^{-1}_{\phi\psi}\langle  V(p_1^{L_0}\phi, 0)V(U(\hat \g)\psi, z)\rangle=\Tr_{\cal H}(p_1^{L_0}),\label{Z1}\ee
which is the standard result for the thermal partition function of a CFT. In the computation, the transformation brings the circles $C_1$ and $C'_1$ to the boundary circles of the annulus around the origin with the radius being $p_1$ and the unit respectively. Now the Schottky generator is simply the diagonal matrix, and $p_1$ is the modular parameter of the torus formed from the annulus by identifying two boundary circles.

More generally we may consider the two-point function of the vertex operators inserted at the fixed points in two circles which are related by an element ${\cL}$ of the Schottky group. As every such element could be put in the form of (\ref{Li}), the two-point function is simply
\be
\langle  ^{\cL}{\bar V V} \rangle = \lim_{z\to \infty}\langle V(U(\hat \g)\phi, z) V(p^{L_0}\phi, 0)\rangle =p^{h},
\ee
where $V$ is the vertex operator corresponding to the state $\phi$ with conformal weight $h$, and $p$ is the multiplier of the  element ${\cL}$. Here we use the notation that the operator $V$ denote the operator inserted  at the fixed point of one circle and $^{\cL}{\bar V}$ is the one in the other circle. Both operators correspond to the same state, with $V$ generating the ket state and $^{\cL}{\bar V}$ generating the bra state.

In the following,  as every Schottky generator is characterized by the fixed points and the multiplier, there is no need to write them explicitly. Formally, the partition function could be written as
\be\label{partition} Z_g\mid_z=\sum_{m_1,m_2,...m_g} \langle
 ~\leftidx{^{{\cL}_1}}{\bar{O}}{^{(1)}_{m_1}}O_{m_1}^{(1)} ~\leftidx{^{{\cL}_2}}{\bar{O}}{^{(2)}_{m_2}}O_{m_2}^{(2)}~~...
  ~\leftidx{^{{\cL}_g}}{\bar{O}}{^{(g)}_{m_g}} ~O_{m_g}^{(g)}\rangle, \ee
where  $m_1,m_2,...m_g$ denote the summation of all of the states on the circles $C_1,C_2,...C_g$ and $C'_1,C'_2,...C'_g$.

\section{Pure AdS$_3$ gravity}

The partition function on a higher genus Riemann surface (\ref{Zg}) could be decomposed into  a summation of $2g$-point correlation functions on Riemann sphere. This is workable for any CFT. It certainly depends on the the spectrum and the OPE of the CFT.  Here we are interested in the large central charge limit of the CFT dual to the AdS$_3$ quantum gravity. In this case, the dual CFT has a sparse light spectrum \cite{Hartman:2013mia,Hartman:2014oaa}, and only the vacuum module
 contributes to  the partition function perturbatively, and other heavy modules  give non-perturbative contribution as $O(e^{-c})$. Therefore we focus on the large central charge limit of the vacuum module. It turns out that the theory becomes essentially free, and the interaction is suppressed in the limit\cite{Fitzpatrick:2014vua,Caputa:2014vaa}.
 After a detailed study of the states in the vacuum module, we compute the genus-1 partition function as a warm up and reproduce the thermal partition function computed in other ways. Next we turn to the computation of the partition function on a higher genus Riemann surface, and  find the perfect agreement with (\ref{1loop}) as well.  

\subsection{Vacuum module in the large c limit}

The vacuum module can be generated by the Virasoro generators acting on the vacuum $\mid 0 \rangle$. The holomorphic sector of the Virasoro algebra is
\be [L_m,L_n]=(m-n)L_{m+n}+\frac{c}{12}m(m^2-1)\delta_{m+n}, \ee
which has a  non-homogenous term of order $c$. The anti-holomorphic sector has the same structure. In the following discussion, we focus on the holographic sector. As the vacuum is invariant under $SL(2,C)$ conformal symmetry, so is annihilated by the generators $L_{\pm1},L_0$. The vacuum module are built on the states
\be\label{state} ...L_{-n}^{r_n}...L_{-3}^{r_3}L_{-2}^{r_2}\mid 0 \rangle, \ee
where only finite number of $r_i$'s are non-zero, and their conformal dimensions are
\be h=\sum_{j=2}^{\inf}jr_j, \ee
in which there is only finite number of non-zero terms in the summation.
A general  state in the module is the linear combination of these states.  We note that the states in (\ref{state}) are not orthogonal to each other.

In the large $c$ limit, the states in the vacuum module could be re-organized more nicely. Under this limit, we can renormalize the operators
\be \hat{L}_m=|\frac{12}{cm(m^2-1)}|^{\frac{1}{2}}L_m ~~~\mbox{for}~|m|\geq 2. \ee
The commutation relations for the renormalized operators are
\bea &~& [\hat{L}_m,\hat{L}_n]=\delta_{m+n}+O(\frac{1}{c^{\frac{1}{2}}}) \notag \\
&~&[L_0,\hat{L}_m]=m\hat{L}_m \notag \\
&~&[L_1,\hat{L}_m]=-sgn(m)|m-1|^{\frac{1}{2}}|m+2|^{\frac{1}{2}} \hat{L}_{m+1} \notag \\
&~&[L_{-1},\hat{L}_m]=-sgn(m)|m+1|^{\frac{1}{2}}|m-2|^{\frac{1}{2}} \hat{L}_{m-1}. \eea
In these   relations, we have absorbed  all of the large $c$ factors into the normalizations of the generators. From the relations, we can read two remarkable facts if we only care about the leading $c$  effects. The first is that the operators $\hat L_m$ and $\hat L_{-m}$ for a fixed $m$ constitute a pair of creation and annihilation operators such that they may build a subspace of the Hilbert space like
\be
\hat L_{-m}^{r_m}|0>, \hs{3ex}\mbox{with $m \in N, m\geq 2$ and $r_m\in N$.}
\ee
Note that the states in different subspace are orthogonal to each other. The other fact is that the state $\hat L_{-m}|0>$ could be constructed by acting
$L_{-1}$ repeatedly for $m-2$ times on the quasiprimary state $\hat L_{-2}|0>=|T>$
\be
\hat L_{-m}|0> \sim (L_{-1})^{m-2}\hat L_{-2}|0>=(L_{-1})^{m-2}|T>.
\ee


A general state in the  vacuum module could be of the form
\be
\prod_{m=2}^{\inf} \hat{L}^{r_m}_{-m} \mid 0 \rangle, \ee
 with only finite number of $r_m$'s being non-zero.  Now different states are orthogonal to each other  to order $c^0$. The normalization for the state is
\be \langle 0\mid \prod_{m=2}^{\inf} \hat{L}^{r_m}_{m} \prod_{m=2}^{\inf} \hat{L}^{r_m}_{-m} \mid 0 \rangle
=\prod_{m=2}^{\inf} r_m!+O(\frac{1}{c}). \ee
 We may define the  ``particle number'' for such a state to be $r=\sum r_m$. The physical reason behind this definition is that each single-particle state $\hat L_{-m}|0>$ corresponds to a graviton.

By contour integral the corresponding operator for the state (\ref{state}) is
\be\label{operator} O_{r_2,r_3,...r_n...}=
:...(\frac{1}{(n-2)!}\partial^{(n-2)}T(z))^{r_n}...(\partial T(z))^{r_3}T(z)^{r_2}:, \ee
which is a product of the stress tensors and their  partial derivatives. It is clear that the ``particle number'' of this operator is the number of the stress tensors  $\displaystyle{r=\sum_{j=2}^{\inf}r_j}$. The two-point function of $O_{r_2,r_3,...r_n...}$ is of order $c^r$ in the large $c$ limit, which means the operator should be normalized with $\frac{1}{c^{{r}/{2}}}$. 

In the following discussion, the single-particle state is of particular importance. For a single-particle state $\hat L_{-m}|0>$, its corresponding vertex operator is of the following forms at the origin and the infinity respectively
\bea V_m &=&(\frac{12}{cm(m^2-1)})^{\frac{1}{2}}\frac{1}{(m-2)!}\partial^{m-2}T(z)\mid_{z=0}, \notag \\
\bar{V}_m &=& (\frac{12}{cm(m^2-1)})^{\frac{1}{2}} \frac{1}{(m-2)!} (-z^2\partial_{z})^{m-2}(z^4 T(z))
\mid_{z\rightarrow \inf} ~~~\mbox{for}~ m\geq2. \eea
At the origin, the normalized vertex operator for the particle-$r$ state (\ref{state}) reads 
\be
\hat O=:(\prod_{j=1}^{r}V_{m_j}):
\ee
In other words, the vertex operator  of a multi-particle state is just the normal ordered product of the vertex operators for the single-particle states.
The important point is that this fact is even true for the states on the circle not around the origin. Under a conformal transformation,  the form of the operator get complicated due to the existence of the partial derivatives. According to (\ref{idenphi}), under a conformal transformation $\g_i$ (\ref{gammai}), the vertex operator at the origin is changed to the one at the fixed point $a_i$
\be
V(\phi_i, 0)\to V(U(\g_i)\phi_i, a_i),
\ee
which could be of a complicated form if $\phi_i$ is a multi-particle state. Taking
\be
\phi_i=L_{-m_1}L_{-m_2}\cdots L_{-m_r}|0>, \hs{3ex}m_i\geq 2, \ee
we find that
\be
U(\g_i)L_{-m_1}L_{-m_2}\cdots L_{-m_r}|0> =U(\g_i)L_{-m_1}U^{-1}(\g_i)U(\g_i)L_{-m_2}U^{-1}(\g_i)\cdots U(\g_i)L_{-m_r}|0>
\ee
Actually the operators $U(\g_i)L_{-m_1}$ and $U(\g_i)L_{-m_1}U^{-1}(\g_i)$ differs only the terms proportional to $L_0$ and $L_{\pm 1}$
\be
U(\g_i)L_{-m_1}=U(\g_i)L_{-m_1}U^{-1}(\g_i)+\mbox{terms involving $L_0$ and $L_{\pm 1}$}.
\ee
As the  states induced by the terms involving $L_0$ and $L_{\pm 1}$ are subdominant in the large central charge limit, we may just
take
\be
U(\g_i)L_{-m_1}L_{-m_2}\cdots L_{-m_r}|0>\sim (U(\g_i)L_{-m_1})(U(\g_i)L_{-m_2})\cdots (U(\g_i)L_{-m_r})|0>.
\ee
In terms of the vertex operators, we have the operator at the fixed point $a_i$ being of the form
\be
V(U(\g_i)\phi_i, a_i)=:\prod_j^r V(U(\g_i)L_{-m_r}|0>, a_i):,
\ee
up to a normalization.
In other words, the vertex operator at $a_i$ could still be written as the normal ordered product of the vertex operators  corresponding to  single-particle states.

In the large $c$ limit, the states constructed above are not only normalized and orthogonal to each other, but also constitute a complete set. Therefore, we may insert such a complete set of states at the pairwise circles in the Riemann sphere to compute the partition function. In other words, we have the relation
\bea\label{identity} \textbf{I}&=&
\mid 0 \rangle \langle 0 \mid +
\sum_{m_1=2}^{\inf} \hat{L}_{-m_1} \mid 0 \rangle \langle 0 \mid \hat{L}_{m_1} +
\frac{1}{2!}\sum_{m_1=2}^{\inf}\sum_{m_2=2}^{\inf} \hat{L}_{-m_1} \hat{L}_{-m_2} \mid 0 \rangle
\langle 0 \mid \hat{L}_{m_2} \hat{L}_{m_1} +... \notag \\
&=&\sum_{r=0}^{\inf}\frac{1}{r!} \sum_{\{m_j\}} (\prod_{j=1}^{r}\hat{L}_{-m_j})
\mid 0 \rangle \langle 0 \mid (\prod_{j=1}^{r}\hat{L}_{m_j})+O(\frac{1}{c}), \eea
where the summation over $m_j$ is from 2 to the infinity, and $r$ is the ``particle number'' for the inserting state.
Here we list the states with the fewest particle numbers in the above relation
\bea\label{identity2} r=0&~~~&\mid 0 \rangle\langle 0 \mid \notag \\
r=1 &~~~&\sum_{m_1=2}^{\inf} \hat{L}_{-m_1} \mid 0\rangle \langle 0 \mid \hat{L}_{m_1} \notag \\
r=2 &~~~&\sum_{2\leq m_1<m_2< \inf } \hat{L}_{-m_1}\hat{L}_{-m_2} \mid 0\rangle
\langle 0 \mid \hat{L}_{m_2} \hat{L}_{m_1} \notag \\
&~~~& \frac{1}{2}\sum_{m_1=2}^{\inf} L_{-m_1}^2 \mid 0 \rangle \langle 0 \mid L_{m_1}^2 \notag \\
...&~~~&
\eea
For a fixed $r$  the state can be written as $\prod\hat{L}_{-j}^{r_j} \mid 0\rangle$, with $\sum r_j =r $, where there is only finite number of $r_j$'s being non-zero. The partition number for the decomposition of $r$ into $\{r_j\}$ is
$\frac{r!}{\prod r_j!}$, that means the state $\prod\hat{L}_{-j}^{r_j} \mid 0\rangle$ will appear $\frac{r!}{\prod r_j!}$ times in the summation in (\ref{identity2}). This cancels the the factor $\frac{1}{r!}$ in (\ref{identity}). The remaining factor is just $1/ \prod r_j!$, which could be cancelled by  the normalization of the state.

The completeness condition could be written in terms of the vertex operators as
 \be \label{identity1} \textbf{I}=\sum_{r=0}^{\inf}\frac{1}{r!} \sum_{\{m_j\}}:(\prod_{j=1}^{r}V_{m_j}):
\mid 0 \rangle \langle 0 \mid :(\prod_{j=1}^{r}\bar{V}_{m_j}):+O(\frac{1}{c^{{1}/{2}}}), \ee
where $::$ denotes the normal ordering. Note that the above complete set of state basis is inserted at the standard unit circle around the origin. Actually on this circle, a ket state  $|A>$ corresponds to the vertex operator $V_A$ inserted at the origin, and a bra state $<A|$ corresponds to the vertex operator $V(U(\hat \g)A)$ inserted at the infinity.

Obviously, one is free to insert the completeness condition at another circle in the Riemann sphere. For example, consider the circle related to the standard circle by the map (\ref{gammai}), the ket state $|A>$ on the circle should created by the vertex operator $V(U(\g_i)A,a_i)$ inserted at the point $a_i$, while the bra state $<A|$ on the circle should created by the vertex operator $V(U(\g_i\hat \g)A,r_i)$ inserted at $r_i$. Inserting an identity operator in the correlation function corresponds to inserting  pairs of the vertex operators at $a_i$ and $r_i$ respectively and summing over all the contribution of the possible vertex operators, i.e.
\be
\langle V_1 V_2 \cdots V_n\rangle =\sum_m \langle V_1 V(U(\g_i)\phi_m,a_i)\rangle \langle V(U(\g_i\hat \g)\phi_m,r_i)V_2 \cdots V_n\rangle. \ee
In particular, if we insert a complete basis  in a two-point function, we have
\be
\langle V_1(z_1) V_2 (z_2)\rangle = \sum_m \langle V_1(z_1) V(U(\g_i)\phi_m,a_i)\rangle \langle V(U(\g_i\hat \g)\phi_m,r_i)V_2 (z_2)\rangle.
\ee
If the operators $V_1,V_2$ correspond to the single-particle states, then among the correlators $\langle V_1(z_1) V(U(\g_i)\phi_m,a_i)\rangle $, the two-point functions between two single-particle states dominate in the large central charge limit so that we have
\be
\langle V_1(z_1) V_2 (z_2)\rangle = \sum_{m=2}^\infty \langle V_1(z_1) V(U(\g_i)L_{-m}|0>,a_i) \rangle \langle V(U(\g_i\hat \g)L_{-m}|0>,r_i)V_2 (z_2)\rangle \label{complete1}
\ee
This relation will play the key role in the following discussion. Note that the relation is true for any $SL(2,C)$ transformation, not only the one in the form (\ref{gammai}).  The above relation could be written schematically as
\be
\langle V_1(z_1) V_2 (z_2)\rangle = \sum_{m=2}^\infty \langle V_1(z_1) \leftidx{^{\cL}}V_{m} \rangle \langle \leftidx{^{\cL}}{\bar V}_{m}V_2 (z_2)\rangle \label{complete2}
\ee
where ${\cL}$ is an $SL(2,C)$ element, $\leftidx{^{\cL}}V_{m} $ is the vertex operator corresponding to the single particle ket state ${\hat L}_{-m}|0>$, and $\leftidx{^{\cL}}{\bar V}_{m}$  corresponds to the bra state. It should be kept in mind that this relation is only true for the vertex operators $V_1,V_2$ corresponding to the single-particle states. The relation (\ref{complete2}) is called the reduced completeness condition. 



As we shown before, the  genus-$g$ partition function could be computed by the $2g$-point functions on the Riemann sphere. In the large central charge limit, these correlation functions are at most of order ${\cO}(c^0)$\cite{Headrick:2015gba}. The order ${\cO}(c^0)$ terms correspond to the 1-loop partition function in the gravity. This could be seen from the operator product expansion(OPE) of the stress tensor:
\be\label{OPE} T(z_1)T(z_2) \sim \frac{{c}/{2}}{(z_1-z_2)^4}+\frac{2T(z_2)}{(z_1-z_2)^2}
+\frac{\partial T(z_2)}{z_1-z_2} +\mbox{normal order}, \ee
and the Ward identity
\bea\label{wald} \langle \prod_{j=1}^{n} T(z_j) \rangle
&=&\sum_{k=2}^{n} \frac{{c}/{2}}{(z_1-z_k)^4}\langle \prod_{\substack{2\leq j \leq n \\ j\neq k}} T(z_j) \rangle
+\frac{2}{(z_1-z_k)^2}\langle  \prod_{2\leq j \leq n} T(z_j) \rangle \notag \\
&~&+\frac{1}{z_1-z_k}\partial_{z_k}\langle  \prod_{2\leq j \leq n} T(z_j) \rangle .
\eea
It is easy to see that the correlation function involving $2n$ stress tensors is at most of order $O(c^{n})$, and the correlation with $2n+1$ stress tensors is at most of order $O(c^{n})$. So the correlation function of the normalized vertex operators  in the Riemann sphere is at most of order $c^0$. More importantly, from the OPE (\ref{OPE}) we see that only the two-point function of the stress tensors is of order $O(c)$ and the three-point function is of order $1$. As a result, we must  focus on the two-point functions of the single-particle states in the large central charge limit. Holographically, this means that we can ignore the interaction of the gravitons, and have a free theory of the gravitons. Therefore, every $2g$-point function in (\ref{ZgV}) could be decomposed into the product of $g$ two-point functions in various ways. The task to compute the partition function (\ref{ZgV}) changes to compute all the possible product of the two-point functions and summing them up. This leads us to prove the 1-loop partition function of a general gravitational configuration.


\subsection{Genus-1 partition function}

Let us first compute the genus-1  partition function in our framework, as a  warm up. In the large $c$ limit, this case has been studied in \cite{Maloney:2007ud}
\be
Z_1=\prod_{m=2}^\infty\frac{1}{1-q^m},
\ee
where $q$ is the modular parameter of the torus. We are now trying to reproduce this result in a new way. Even though the derivation looks tedious, the computation is suggestive for the computation in  higher genus cases.

In the torus case, the Schottky group is generated by only  one $SL(2,C)$ element ${\cL}$.
The genus-1 partition function could be read from
\bea\label{genus1} Z_1
&=&\sum_{r=0}^{\inf}\frac{1}{r!} \sum_{\{m_j\}} \langle  :(\prod_{j=1}^{r}\leftidx{^{\cL}}{\bar{V}}_m(r_1)):~
:(\prod_{j=1}^{r}V_m(r_1)):  \rangle +O(\frac{1}{c}),
\eea
By the OPE  (\ref{OPE}) and the Ward identity (\ref{wald}), the expectation value for $2r$ stress tensors are at most order $c^r$ and the leading $c$ contribution are captured by  the products of $r$ two-point functions, and the partition function is the summation of all the products with appropriate combinatory factors. For $r=0$ term,  the contribution from the vacuum is $1$.
For $r=1$ term
\be\label{r1} Z^{(1)}=\sum_{m=2}^{\inf}\langle \leftidx{^{\cL}}{\bar{V}}_m(r_1) V_m(a_1)\rangle
=\Tr_{{\cal H}_1}q^{L_0}
=\sum_{m=2}^{\inf} q^m,
\ee
where we have used the relation (\ref{Z1}) but now only sum the single particle states ${\cal H}_1$. 
For $r>1$ case, the expectation value equals to
\bea\label{generalr} &~&\frac{1}{r!}\sum_{m_1=2}^{\inf}\sum_{m_2=2}^{\inf}...\sum_{m_r=2}^{\inf} \langle :\leftidx{^{\cL}}{\bar{V}}_{m_1}(r_1)\leftidx{^{\cL}}{\bar{V}}_{m_2}(r_1)...\leftidx{^{\cL}}{\bar{V}}_{m_r}(r_1):
:V_{m_1}(a_1)V_{m_2}(a_1)...V_{m_r}(a_1): \rangle \notag \\
&=&\frac{1}{r!}\sum_{m_1=2}^{\inf}\sum_{m_2=2}^{\inf}...\sum_{m_r=2}^{\inf}
\sum_{\{P\}} \langle  \leftidx{^{\cL}}{\bar{V}}_{m_{P_1}}(r_1)V_{m_1}(a_1)\rangle
\langle  \leftidx{^{\cL}}{\bar{V}}_{m_{P_2}}(r_1)V_{m_2}(a_1)\rangle...
\langle \leftidx{^{\cL}}{\bar{V}}_{m_{P_r}}(r_1)V_{m_r} (a_1)\rangle +O(c^{-1}),\label{rsum}
 \eea
 where $P$ denote different permutation. There is no two-point function between two $V$ operators or two $\bar{V}$ operators at the same fixed point because of normal ordering.

\begin{figure}
  \centering
  \includegraphics[width=8cm]{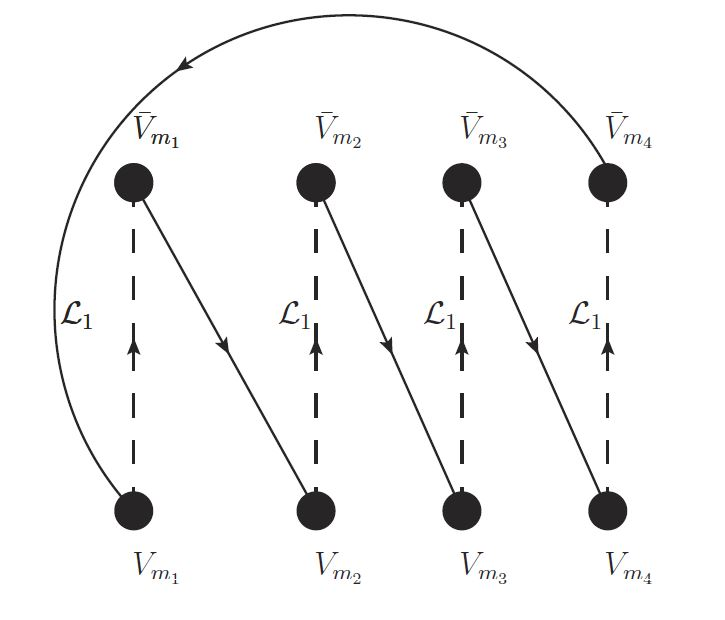}\\
  \caption{The link formed by the product of four two-point functions. It corresponds to the conjugacy class ${\cal L}^4$. Due to the normal ordering, the only possible connected link is the one in the diagram. }\label{multi}
\end{figure}

 To classify the possible combination of two-point functions in the summation (\ref{rsum}) clearly, we define a diagram language. As in Fig.\ref{multi}, the dotted vertices denote the fixed points, where the operators are inserted: the lower ones are the $V_m(a_1)$'s, while  the upper ones are the $\leftidx{^{\cL}}{\bar{V}}_m(r_1)$'s. The dashed lines denote the summations over $m_i$'s and the solid line denotes the correlation between two vertex operators. The dashed and solid lines may form a closed contour, which will  be called as a link. In short, a link is defined by certain product of two-point function of single-particle operators. The length of the link is defined to be the  number of dashed lines. It is convenient to assign a direction on the dashed line indicating the flow between $V$ to $\bar{V}$. As we will see shortly, the direction from $V$ to $\bar{V}$ corresponds to a Schottky generator, while the flipped direction corresponds to the inverse of the generator. The expectation value of a link is
\bea\label{link} &~&\sum_{\{m_t\}}\langle  \leftidx{^{\cL}}{\bar{V}}_{m_{t_2}} V_{m_{t_1}}\rangle
\langle  \leftidx{^{\cL}}{\bar{V}}_{m_{t_3}}V_{m_{t_2}}\rangle ~...
~\langle  \leftidx{^{\cL}}{\bar{V}}_{m_{t_1}}V_{m_{t_s}}\rangle \notag \\
&=& \sum_{m_{t_1}=2}^{\inf}\sum_{m_{t_2}=2}^{\inf}...\sum_{m_{t_s}=2}^{\inf}
 \langle  \leftidx{^{\cL}}{\bar{V}}_{m_{t_2}} V_{m_{t_1}}\rangle
\langle  \leftidx{^{{\cL}^2}}{\bar{V}}_{m_{t_3}}\leftidx{^{\cL}}{V}_{m_{t_2}}\rangle ~...
~\langle  \leftidx{^{\cL}}{\bar{V}}_{m_{t_1}}V_{m_{t_s}}\rangle  \notag \\
&=& \sum_{m_{t_1}=2}^{\inf}\sum_{m_{t_3}=2}^{\inf}...\sum_{m_{t_s}=2}^{\inf}
 \langle  \leftidx{^{{\cL}^2}}{\bar{V}}_{m_{t_3}}V_{m_{t_1}} \rangle
 \langle \leftidx{^{{\cL}}}{\bar{V}}_{m_{t_4}} {V}_{m_{t_3}}\rangle ~...
~\langle  \leftidx{^{\cL}}{\bar{V}}_{m_{t_1}}V_{m_{t_s}}\rangle  \notag \\
&=& \sum_{m_{t_1}=2}^{\inf} \langle  \leftidx{^{{\cL}^{s}}}{\bar{V}}_{m_{t_1}}V_{m_{t_1}} \rangle
=\sum_{m=2}^{\inf} q^{ms}
\eea
where $s$ is the length of the link. Here we have used the fact that the two-point functions are invariant under conformal transformation and the reduced completeness condition (\ref{complete2}) for the two-point function. The above symbolic derivation could be made more clear with the explicit expressions of the vertex operators. As ${\cL}=\g_1 \g_q \g^{-1}_1$, ${\cL}^s=\g_1 (\g_q)^s \g^{-1}_1$,  and the computation of the two-point functions in the last step is on an annulus with   the modular parameter being $q^s$.

The two-point function of the particle-$r$  vertex operators can be cast into  the summation of the products of $r$ two-point functions of single-particle operators. Each product can be decomposed into several links, with the contribution of each link being (\ref{link}). Assuming the $r$ pairs of the operators can be decomposed into a series of links  such that
\be \sum_{s=1}^{\inf} s t_s=r, \ee
where $t_s$ is the number of the length-$s$ links.
The partition number for this decomposition is $\frac{r!}{\prod_{s=1}^{\inf} (s!)^{t_s}t_s!}$. Furthermore for each patch with $s$ pairs of the operators there are $(s-1)!$ different ways to build the connected link, so the combinatory factor is $\frac{r!}{\prod_{s=1}^{\inf} s^{t_s}t_s!}$. The numerator is cancelled by  the coefficient $\frac{1}{r!}$ in the partition function so the overall coefficient is $\frac{1}{\prod_{s=1}^{\inf} s^{t_s}t_s!}$, which is remarkably independent of $r$.

For the partition function (\ref{genus1}), we just need to sum over all the contributions from  different combinations of the links
\bea Z_1&=&\prod_{s=1}^{\inf}\sum_{t=0}^{\inf}\frac{1}{s^{t}}\frac{1}{t!}(\sum_{r=2}^{\inf}q^{sr})^t
=\exp \sum_{r=2}^{\inf}- \log (1-q^r)=\prod_{r=2}^{\inf} \frac{1}{1-q^r}. \eea
This is the genus-1 partition function found in \cite{Maloney:2007ud}. It is actually the one for the holomorphic sector, and there is the anti-holomorphic sector which gives the same contribution. On the other hand, in the gravitational partition function, there is the contribution from the primitive conjugacy class ${\cL}^{-1}$, which is the same as the one from ${\cL}$.

 From our derivation, it seems to be of order $O(c^0)$ but it in fact is
exact without higher order $1/c$ correction. The exactness of the genus-1 partition function could be seen from the relation (\ref{Z1}), which shows that the function depends only on the spectrum of the vacuum module.

\subsection{Genus-2}

\begin{figure}
  \centering
  \includegraphics[width=8cm]{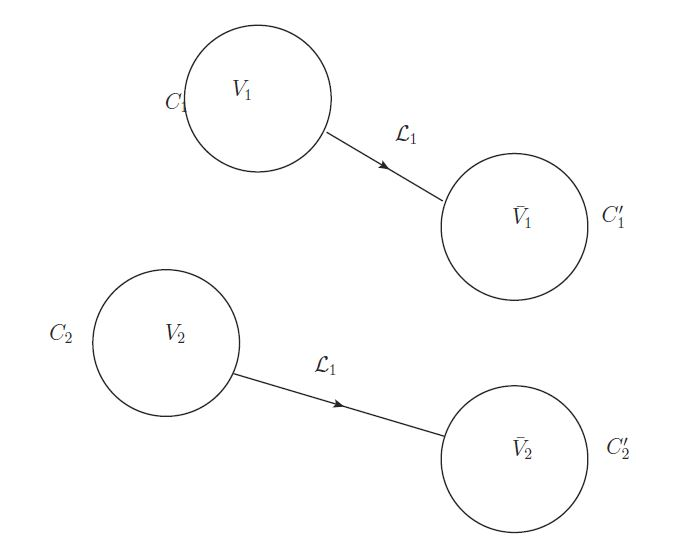}\\
  \caption{The circles correspond to the genus-2 Riemann surface}\label{genus2}
\end{figure}

\begin{figure}[tbp]
  \centering
  \subfloat[The link corresponds to ${\cL}_1$ and ${\cL}_2$.]{\includegraphics[width=6cm]{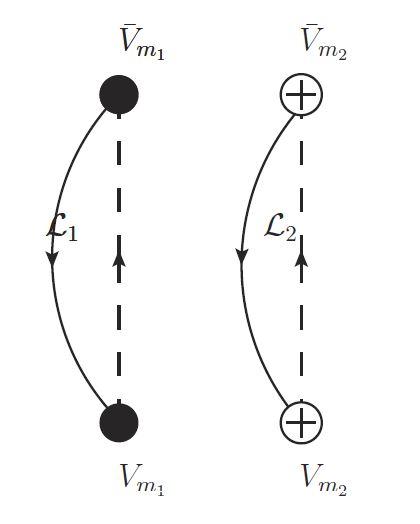}
  \label{2point1}}
   \subfloat[The link corresponds to ${\cL}_1{\cL}_2$.] {\includegraphics[width=7cm]{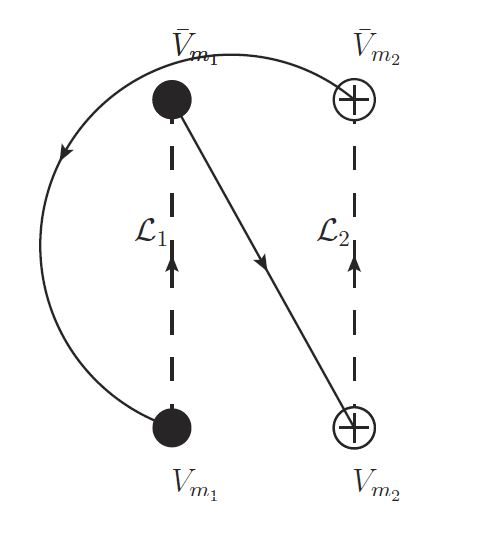}
  \label{2point2}}\\
  \caption{In the diagram, the same type of vertices means that the operators are in the fixed points of the pairwise circles in the Schottky uniformization. The two-point function between the operators on the same type of vertices just give the simplest link. The one between the operators on different types of vertices may lead to more complicated links.}
  \end{figure}

For the genus 2 case, there are two free generators in the Schottky group. In the Riemann sphere, there are four circles with the fixed points $a_1,a_2$ and $r_1,r_2$, as shown in Fig. \ref{genus2}. The partition function could be written as
 \be
Z_2=\sum_{m_1,m_2}\langle ~ ^{{\cL}_1}{\bar O}^{(1)}_{m_1}O^{(1)}_{m_1} ~ ^{{\cL}_2}{\bar O}^{(2)}_{m_2} O^{(2)}_{m_2}\rangle
\ee
where $m_1,m_2$ are over all possible states in the vacuum module. For the multi-particle states, every operator $O_{m_i}$ could be decomposed into the product of the operators corresponding to the single-particle states. To simplify the discussion, let us first consider the simplest case that four operators in the correlation functions are all single-particle operators. As in genus-1 case, the four-point correlator could be decomposed into the product of two two-point functions. However, there are now more possibility for the operators to combine. For example, the vertex operator at $a_1$ can not only contract with the operator at $r_1$, but can also contract with the operators at $a_2$ and $r_2$. Without losing generality, we assume the operators at $a_1,r_1$ correspond to the state ${\hat L}_{-m_1}|0>$, and the operators at $a_2,r_2$ correspond to the state ${\hat L}_{-m_2}|0>$.  In the first case, when the operator at $a_i$ is connected with the one at $r_i$ to form two-point functions as in Fig. \ref{2point1}, it is easy to see that the contribution is simply
 \be
\sum_{m_1,m_2}\langle \leftidx{^{{\cL}_1}}{\bar{V}}_{m_1}(r_1) V_{m_1}(a_1)\rangle \langle \leftidx{^{{\cL}_2}}{\bar{V}}_{m_2}(r_2) V_{m_2}(a_2)\rangle=\sum_{m_1}(q_1)^{m_1}\sum_{m_2}(q_2)^{m_2}
\ee
where $q_1$ and $q_2$ are respectively the multipliers in the generators ${\cL}_1$ and ${\cL}_2$. If we consider all the states but only allow the operators connect from $a_i$ to $r_i$, then finally we get the product of two genus-1 partition functions $Z_1(q_1)Z_1(q_2)$.

On the other hand, we are allowed to connect the operators at $a_1$ to the one at $a_2$ or $r_2$. Let us first consider the case that the operator at $a_1$ connect to the one at $r_2$  as in Fig. \ref{2point2}, then the contribution could be
\bea
\lefteqn{\sum_{m_1,m_2}\langle \leftidx{^{{\cL}_2}}{\bar{V}}_{m_2}(r_2) V_{m_1}(a_1)\rangle \langle \leftidx{^{{\cL}_1}}{\bar{V}}_{m_1}(r_1) V_{m_2}(a_2)\rangle}\nn\\
&=&\sum_{m_1,m_2}\langle \leftidx{^{{\cL}_1{\cL}_2}}{\bar{V}}_{m_2} \leftidx{^{{\cL}_1}}V_{m_1}\rangle \langle \leftidx{^{{\cL}_1}}{\bar{V}}_{m_1} V_{m_2}\rangle\nn\\
&=&\sum_{m_2}\langle \leftidx{^{{\cL}_1{\cL}_2}}{\bar{V}}_{m_2} V_{m_2}\rangle \nn\\
&=&\sum_{m}(q_{12})^{m}
\eea
where $q_{12}$ is the multiplier of the $SL(2,C)$ element ${\cL}_1{\cL}_2$. Here we have used the completeness condition (\ref{complete2}) and the fact that the two-point function is conformal invariant. Certainly we may make conformal transformation on the second two-point function and use the completeness condition on $V_{m_2}$, and find
\bea
\sum_{m_1,m_2}\langle \leftidx{^{{\cL}_2}}{\bar{V}}_{m_2}(r_2) V_{m_1}(a_1)\rangle \langle \leftidx{^{{\cL}_1}}{\bar{V}}_{m_1}(r_1) V_{m_2}(a_2)\rangle
=\sum_{m_1}\langle ^{{\cL}_2{\cL}_1}{\bar{V}}_{m_1}V_{m_1}\rangle=\sum_{m}(q_{21})^{m},
\eea
where $q_{21}$ is the multiplier of the element ${\cL}_2{\cL}_1$. However, note that the element ${\cL}_2{\cL}_1$ is in the same primitive conjugacy class as ${\cL}_1{\cL}_2$, and have the same multiplier so that $q_{12}=q_{21}$.

\begin{figure}
  \centering
  \subfloat[The link corresponds to ${\cL}_1{\cL}^{-1}_2$.]
  {\includegraphics[width=6cm]{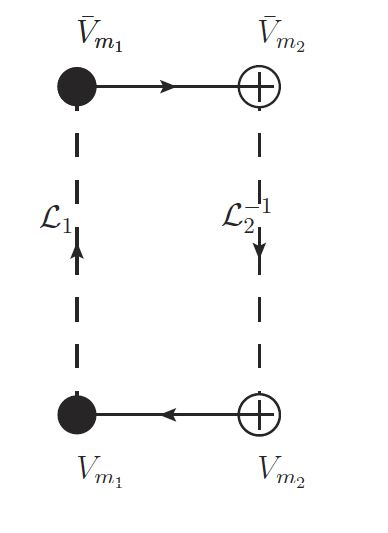}\label{2point31} }
  \subfloat[The link corresponds to ${\cL}_2{\cL}^{-1}_1$.]
  {\includegraphics[width=6cm]{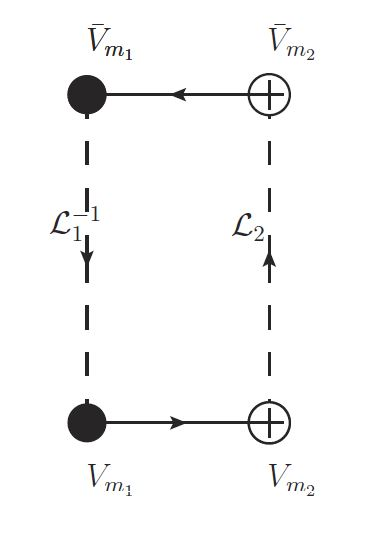}\label{2point32}}\\
  \caption{Two links with opposite orientations. The corresponding conjugacy classes are inverse to each other. }
  \end{figure}

For the contraction that the operator  at $a_1$ connect to the one at $a_2$  as in Fig. \ref{2point31}, the contribution is
\bea
\sum_{m_1,m_2}\langle \leftidx{^{{\cL}_2}}{\bar{V}}_{m_2}(r_2) \leftidx{^{{\cL}_1}}{\bar{V}}_{m_1}(r_1)\rangle \langle  V_{m_1}(a_1)V_{m_2}(a_2)\rangle&=&\sum_{m_2}\langle ^{{\cL}_2}{\bar{V}}_{m_2} ^{{\cL}_1}V_{m_2}\rangle=\sum_{m}(q_{\bar{1}2})^{m} \label{2point31q}
\eea
where $q_{\bar{1}2}$ is the multiplier of the element ${\cL}^{-1}_1{\cL}_2$. Note that the element ${\cL}^{-1}_1{\cL}_2$ belongs to a different conjugacy class from ${\cL}_1{\cL}_2$. However, it could also give
\bea
\sum_{m_1,m_2}\langle \leftidx{^{{\cL}_2}}{\bar{V}}_{m_2}(r_2) \leftidx{^{{\cL}_1}}{\bar{V}}_{m_1}(r_1)\rangle \langle  V_{m_1}(a_1)V_{m_2}(a_2)\rangle&=&\sum_{m_1}\langle ^{{\cL}_1}{\bar{V}}_{m_1} ^{{\cL}_2}V_{m_1}\rangle=\sum_{m}(q_{\bar{2}1})^{m} \label{2point32q}
\eea
where $q_{\bar{2}1}$ is the multiplier of the element ${\cL}^{-1}_2{\cL}_1$. Now there appears another conjugacy class ${\cL}^{-1}_2{\cL}_1$ which is the inverse of the class  ${\cL}^{-1}_1{\cL}_2$. Fortunately, as the multipliers of an element and its inverse are the same, both ways lead to the same answer.

It is easy to understand the conjugacy class from the diagram with  directions on the dashed lines. In the first case that $a_i$ connect with $r_i$, the dashed lines and the solid lines starting and ending at the same points form two links, each of which corresponds to the primitive element ${\cL}_i$, see Fig. \ref{2point1}. In the second case that $a_1$ ($a_2$) connect to $r_2$ ($r_1$), the arrows along the dashed lines in the link point to the same direction, this correspond to the conjugate element ${\cL}_1{\cL}_2$ , see Fig. \ref{2point2}.  In the last case that $a_i$ and $b_i$ form a square, the directions on two dashed lines are opposite, suggesting the corresponding element is ${\cL}_1{\cL}^{-1}_2$ or ${\cL}_2{\cL}^{-1}_1$, depending on which vertex we start from, see Fig. \ref{2point31} and \ref{2point32}. Note that  from the diagram, if the orientation of the link gets inverted, then the  corresponding  conjugacy class is inverse to the original one.  For example, if we change the orientation in Fig. \ref{2point2}, we find the conjugacy class ${\cL}_1^{-1}{\cL}_2^{-1}$.
In Fig. \ref{2point1} and Fig. \ref{2point2}, the  reversing of the orientation would not happen, as there is only one kind of orientation no matter where the starting point of the link is. However, in Fig. \ref{2point31} and \ref{2point32}, there could be two opposite orientations depending on the starting point. This is the reason why we get two different relations (\ref{2point31q}) and (\ref{2point32q}). The difference depends on which  two-point function appears in the last step, the one between $\bar{V}_{m_1}$ and $V_{m_1}$, or the one between   $\bar{V}_{m_2}$ and $V_{m_2}$.  In other words,
 the computation of the correlation function cannot distinguish the difference between two conjugacy elements which are inverse to each other. Nevertheless, the different ways in computing lead to the same result.

If the operators correspond to the multi-particle states, then the situation becomes complicated. Now we are allowed to form link not only between the operators at $a_1$ and $r_1$, but also between the operators at $a_1$ and $a_2$ or $r_2$ at the same time. The different ways of forming the loop lead to different conjugate class. In the next subsection, we will have a systematic discussion for higher genus case.

\subsection{Higher genus}

In a general higher genus case, we can compute the partition function by inserting the states in the vacuum module  at each circle. This leads to compute the $2g$-point functions in (\ref{partition}). The states inserting at the different circles include the states with various particle numbers. We assume that the particle number of the states at $C_i$ to be $r_i$, then the corresponding contribution to the partition function is
\bea\label{highergenus} Z_{r_1,r_2,...r_g}&=&
\prod_{t_1=1}^{r_1} (\sum_{m_{1,t_1}=2}^{\inf})
\prod_{t_2=1}^{r_2} (\sum_{m_{2,t_2}=2}^{\inf})...
\prod_{t_g=1}^{r_g} (\sum_{m_{g,t_g}=2}^{\inf})
\langle \frac{1}{r_1!}(:\prod_{t_1=1}^{r_1} \leftidx{^{{\cL}_1}}{\bar{V}}^{(1)}_{m_{1,t_1}}:)
(:\prod_{t_1=1}^{r_1} V^{(1)}_{m_{1,t_1}}:) \notag \\
&~& \cdot \frac{1}{r_2!}(:\prod_{t_2=1}^{r_2} \leftidx{^{{\cL}_2}}{\bar{V}}^{(2)}_{m_{2,t_2}}:)
(:\prod_{t_2=1}^{r_2} V^{(2)}_{m_{2,t_2}}:)...
\frac{1}{r_g!}(:\prod_{t_g=1}^{r_g} \leftidx{^{{\cL}_g}}{\bar{V}}^{(g)}_{g,m_{t_g}}:)
(:\prod_{t_g=1}^{r_g} V^{(g)}_{g,m_{t_g}}:)  \rangle. \notag 
\eea
Here we take the notation that in $V^{(i)}_{m_{i,t_i}}$ the $i$ labels the circle, $t_i$ denotes the particle index and every $m_{i,t_i}$ takes value from $2$ to $\infty$.

\begin{figure}
  \centering
  \subfloat[The link corresponds to ${\cL}^2_1{\cL}^{-1}_2{\cL}_3$.]
  {\includegraphics[width=6cm]{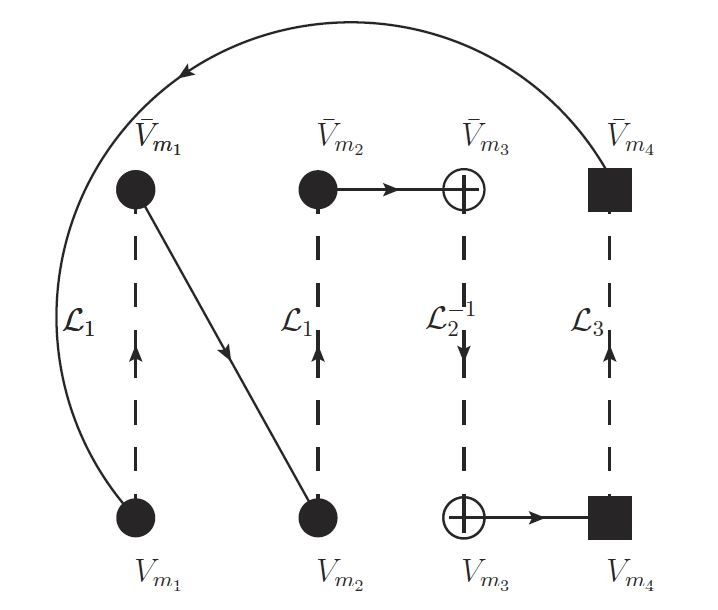}\label{3point0} }
  \subfloat[The link corresponds to ${\cL}^{-1}_3{\cL}_2{\cL}^{-2}_1$.]
  {\includegraphics[width=6cm]{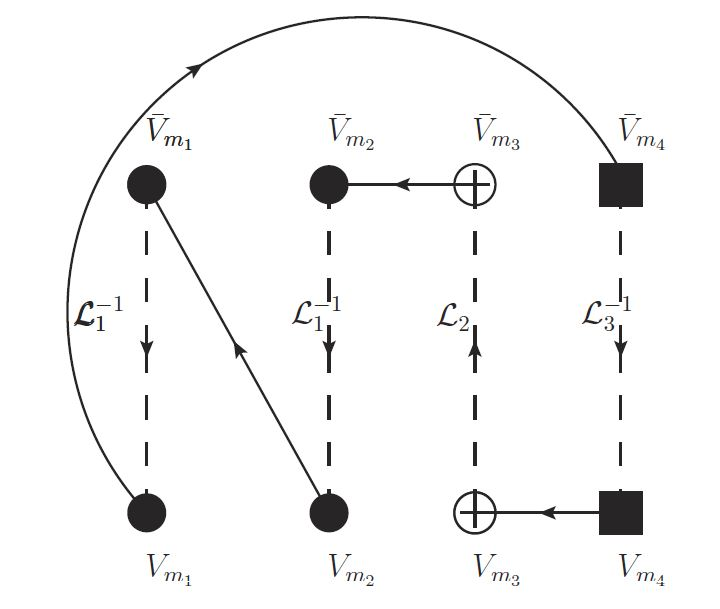}\label{3point1}}\\
  \caption{More complicated links with three generators.  }
  \end{figure}

As shown in the genus-2 case, the operators inserted  at the fixed point in the circle $C_i$ are free to contract with the operators in other circles, including $C'_i$. It is convenient to use the diagrammatic language introduced above to characterize all the possible contractions. A general contraction between the operators can form a closed link by the dashed lines and solid lines.  The linking number is the number of the dashed lines, labelled by the circles and its related Schottky generators. We may use the symbol to characterize a link as
 \be\label{long} \hat l=(j_1^{r_1}j_2^{r_2}...j_s^{r_s}), \ee
where $j_k^{r_k}$ means that there are $|r_k|$ pairs of operators inside the circles $C_k$ and $C'_k$ appearing continually in the correlator such that it contributes a linking number $|r_k|$. The value of $r_k$ could be positive or negative, up to the flow is from $V$ to $\bar V$ or vice versa. Note that in a link, the flow should be continuous. For example, in the genus-$1$ case, the link between two single particle state is just a link $(1^1)$, while the one formed by the particle-$k$ states is $(1^k)$. In the genus-$2$ case, the link in Fig.\ref{2point1} is just $(1^1)(2^1)$, the one in Fig.\ref{2point2} is $(1^12^1)$, while the one in Fig.\ref{2point31} is $(1^12^{-1})$ and the one in Fig. \ref{2point32} is $(1^{-1}2^1)$.  One subtle point is that for one link diagram, there could be two different orientation, like the ones in Fig. \ref{3point0} and Fig.\ref{3point1}, with the corresponding  conjugacy classes being inverse to each other.   In more general case, the operators in one pair of circles may appear in different positions of a link, and it is forbidden to permute their positions if there are other operators between them. Namely, there could be a link of the form
\be
(1^{k_1}2^{k_2}1^{k_3}2^{k_4}),
\ee
 which is different from the link $(1^{k_1+k_3}2^{k_2+k_4})$.

 More importantly, an oriented link is in one-to-one correspondence with the conjugacy class of the Schottky group. The element corresponding to the link (\ref{long}) is
 \be
 {\cL}_{j_1}^{r_1}{\cL}_{j_2}^{r_2}\cdots {\cL}_{j_s}^{r_s}.
 \ee
 Note that an oriented link has a cyclic symmetry, as there is  freedom to start labelling a link from any point on the link. Remarkably, the different elements corresponding to the different labels are conjugate to each other.
 Recall that in a free generated group, the group elements could be formed from the generators and their inverses. If we take the generators and their inverses as the alphabets, we can construct ``words" with the letters. If a generator and its inverse appear next to each other in a word, the word could be simplified by omitting these two letters.  A reduced word is the word which cannot be simplified. Moreover, a word is called cyclically reduced if its first and last letters are not inverse to each other. Every reduced word is conjugate to a cyclically reduced word, and a cyclically reduced conjugate of a cyclically reduced word is a cyclic permutation of the letters in the word. Simply speaking, a cyclically reduced conjugate stand for a conjugacy class. As in our discussion, the vertex operators have been normal ordered so there is no contraction in the same vertex operators. This in fact forbids the appearance of the generators ${\cL}_i {\cL}_i^{-1}$ in a link. Namely, the simplification and the cyclically reduction in the word  have been encoded in the definition of the vertex operators.
   Therefore one oriented link corresponds to one conjugacy class.
 A primitive conjugate element is the one which cannot be written as the positive power of another element, i.e. ${\cL}^{(primary)} \neq ({\cL}')^n, n\in N$.
 It corresponds to the link which cannot be written as the positive powers of a shorter link. For example, the link $(1^k)$ is not primitive, as $(1^k)=(1)^k$.

It is remarkable that the link formed from the product of the two-point function is generically of two opposite orientations,  corresponding to two conjugacy classes inverse to each other. However,  both conjugacy classes have the same multiplier and therefore the correlation function gets the same value. In the following, we ignore the orientation of the link, but keep in mind that the conjugacy classes have been doubly counted.

 The value of a link is easy to compute. For a link corresponding to a primitive class, its value is just
 \be
 Z_{\hat{l}}=\sum_{m=2}^{\inf} (q_{\hat{l}})^m,
 \ee
 where $q_{\hat l}$ is the  multiplier of the primitive element corresponding to the link. For a non-primitive link which can be written as ${\hat l}=({\hat l}_{(p)})^n$, its value is
 \be
 Z_{\hat l}=\sum_{m=2}^\infty (q_{{\hat l}_{(p)}})^{nm}, \ee
where $q_{{\hat l}_{(p)}}$ is the multiplier of the primitive element corresponding to the link ${\hat l}_{(p)}$.

A general $2g$-point function of multi-particle operators on a Riemann sphere can be decomposed into the summation of the product of the  links. One kind of link can appear multiple times in the product.  The order of the links does not matter, and one can move the links freely. The  important thing is the coefficient for the multiple links. For a diagram with $r_1, r_2,..., r_g$ particles at the circles respectively, the permutation  among the particles gives the same kind of diagram. The permutation at the circle $C_j$ contributes a   $r_j!$ factor, which cancel the coefficient in (\ref{highergenus}). However such permutation has two kinds of degeneracy.
\begin{enumerate}
\item If the link $\hat l$ which have the linking number $l$ appears $n_l$ times, then when we permute these links we get the same diagram.  There is a $n_l!$ degeneracy of over-counting in this case.
\item If there is a link which is not primitive, being the $s$-th power of a primitive link, then there is a translational symmetry along the primitive element. This symmetry contributes an order-$s$ degeneracy.
\end{enumerate}
Therefore for a non-primitive link appearing $n_l$ times in the product, their contribution is
\be
Z_{n_l}=\left(\frac{1}{s}\right)^{n_l}\frac{1}{n_l!}\left(\sum_{m=2}^\infty q_{{\hat l}_{(p)}}^{sm}\right)^{n_l}, \label{nl}
\ee
where $q_{{\hat l}_{(p)}}$ is the multiplier of the primitive element in the link.

Now we are ready to show the equivalence between the $2g$-point function (\ref{ZgV}) on the Riemann sphere and the 1-loop partition function (\ref{1loop}). First of all,  there is an one-to-one correspondence between the primitive link and primitive conjugacy class in the Schottky group. By considering all possible links, there is no missing in counting every primitive element. Moreover, notice that the 1-loop partition function (\ref{1loop})
could be expanded
\be
Z_{1-loop}=\prod_\g  Z_\g=\prod_\g\left(\prod_{m=2}^\infty \frac{1}{1-q_\g^m}\right),
\ee
and the contribution from each primitive element could be expanded further
\be
\prod_{m=2}^\infty \frac{1}{1-q_\g^m}=\sum_{t=0}^\infty \frac{1}{t!}\prod_{s=1}^\infty \frac{1}{s^t}(\sum_{m=2}^\infty q_\g^{sm})^t.
\ee
Compared to (\ref{nl}), It is obvious that each term in the summation is the contribution of a kind of link  which appear $t$ times. This kind of link could be non-primitive. Therefore,  the 1-loop partition function could be expanded into a summation of the contribution from all possible links, resulted from the contraction of two-point functions in the $2g$-point function (\ref{ZgV}).  This proves that the 1-loop partition function (\ref{Zg}) is captured exactly by the $2g$-point function (\ref{ZgV}) in the large central charge limit.

In the above discussion, we have been focusing on the holomorphic sector the CFT. The anti-holomorphic sector should give the same contribution.
This requires us to take the square of the result in the holomorphic sector, and may bring mismatch with the gravitational 1-loop result. However, the computation in the CFT cannot distinguish the link with different orientation, though we may set up the one-to-one correspondence between the oriented links and conjugacy classes.   On the other hand, in computing (\ref{1loop}), $q^{-1/2}_\g$ should be the larger values of the element $\g$ so that it is actually the same for both $\g$ and $\g^{-1}$.  Therefore a more precise relation is
\be Z_g|_{holomorphic}=\prod_{\gamma}(Z_\gamma)^{\frac{1}{2}} \ee
But the full partition function including both holomorphic and anti-holomophic sector indeed match with (\ref{1loop})
\be Z_g=\prod_{\gamma}|Z_\gamma| .\ee

\section{Higher spin}

In previous section, we calculate the higher-genus partition function for the theory dual to pure AdS$_3$ gravity at the order $c^0$ in the large $c$ limit. In this section, we extend the study to the higher spin AdS$_3$ gravity and its CFT dual. As in the pure AdS$_3$ gravity, one may rewrite the action of the higher spin AdS$_3$ gravity in terms of two decoupled Chern-Simons actions with opposite levels. The gauge group could be enlarged from $SL(2,R)$ to $SL(n,R)$ in order to describe the higher spin fields up to spin $n$. It was found that by imposing generalized Brown-Henneaux boundary condition, the asymptotic symmetry group of higher spin gravity is ${\cal W}$ symmetry with the central charge $c=\frac{3l}{2G}$\cite{Campoleoni:2010zq, Henneaux:2010xg}. This indicates that the higher spin AdS$_3$ gravity is dual to 2D CFT with ${\cal W}$ symmetry. Here we only focus on the semiclassical higher spin AdS$_3$ gravity and study the handlebody configurations which could be obtained as the quotient of the  global AdS$_3$ by a Schottkky group. In these cases, the bulk configurations are the same as the ones in pure AdS$_3$ gravity, but the fluctuations around these configurations must include the higher spin ones, besides the usual spin $2$ graviton. The 1-loop partition function now turns out to be
\be
Z_{1-loop}=\prod_{s=2} Z_{1-loop,s}=\prod_{s=2}\prod_\g\left(\prod_{m=s}^\infty \frac{1}{1-q_\g^m}\right)\label{1loops}
\ee
where the contribution from the fluctuation of different spins could be factorized
\be
Z_{1-loop,s}=\prod_\g\left(\prod_{m=s}^\infty \frac{1}{1-q_\g^m}\right).
\ee

On the dual side, the conformal field theory has an ${\cal W}$ symmetry.
In the large $c$ limit, the algebra is simplified to be
\bea &~&[\hat{L}_m,\hat{L}_n]=\delta_{m+n} \notag \\
&~&[\hat{W}_m,\hat{W}_n]=\delta_{m+n} \notag \\
&~&[\hat{L}_m,\hat{W}_n]=0, \eea
where
\bea &~&\hat{L}_m=(\frac{12}{|m(m^2-1)c|})^{\frac{1}{2}}L_m ~~~\mbox{for}~|m| \geq 2 \notag \\
&~&\hat{W}_m=(\frac{36}{|-\sigma m(m^2-1)(m^2-4)c|})^{\frac{1}{2}}W_m~~~\mbox{for}~|m| \geq 3. \eea
Thus in a CFT with ${\cal W}$ symmetry one has to consider not only the vacuum module consisting of the states generated by $L_{-m}, m\geq 2$, but also the modules generated by the $W$ primaries and their descendants, in the large $c$ limit.
 In the module generated by a  $W$ primary, the lowest-weight state is generated by the primary field. For example, in the CFT with ${\cal W}_3$ symmetry, the lowest-weight  state in the $W_3$ module is the one generated by the $W_{-3}|0>$, and the other states could be obtained by acting $L_{-1}$ repeatedly on $W_{-3}|0>$.  This is very similar to the construction in the vacuum module, where the lowest weight state is generated by $L_{-2}$. Correspondingly, the holomorphic vertex operators in the $W_3$ module are of the forms
 \be
 W_3(z), \hs{2ex}\p W_3, \hs{2ex}\p^2 W_3 \cdots .
 \ee
 For the CFT with other ${\cal W}$ symmetries, the discussion is similar. One has to keep in mind that in the large $c$ limit, the modules generated by  different primaries are decoupled. The completeness condition for the CFT with ${\cal W}_3$ symmetry now changes to
 \bea\label{identityW3} \textbf{I}&=&
\mid 0 \rangle \langle 0 \mid +
\sum_{m_1=2}^{\inf} \hat{L}_{-m_1} \mid 0 \rangle \langle 0 \mid \hat{L}_{m_1} +
\frac{1}{2!}\sum_{m_1=2}^{\inf}\sum_{m_2=2}^{\inf} \hat{L}_{-m_1} \hat{L}_{-m_2} \mid 0 \rangle
\langle 0 \mid \hat{L}_{m_2} \hat{L}_{m_1} +... \notag \\
&&+
\sum_{n_1=3}^{\inf} \hat{W}_{-n_1} \mid 0 \rangle \langle 0 \mid \hat{W}_{n_1} +
\frac{1}{2!}\sum_{n_1=3}^{\inf}\sum_{n_2=3}^{\inf} \hat{W}_{-n_1} \hat{W}_{-n_2} \mid 0 \rangle
\langle 0 \mid \hat{W}_{n_2} \hat{W}_{n_1} +... \notag \\
& &+\frac{1}{2!}\sum_{m_1=2}^{\inf}\sum_{n_2=3}^{\inf} \hat{L}_{-m_1} \hat{W}_{-n_2} \mid 0 \rangle \langle 0 \mid \hat{W}_{n_2} \hat{L}_{m_1} +... \notag \\
&=&\sum_{r=0}^{\inf}\frac{1}{r!} \left(\sum_{\{m_j\}\{n_k\}} (\prod_{j=1}^{r_1}\hat{L}_{-m_j})(\prod_{k=1}^{r-r_1}\hat{W}_{-n_k})\mid 0 \rangle \langle 0 \mid (\prod_{j=1}^{r_1}\hat{L}_{m_j})(\prod_{k=1}^{r-r_1}\hat{W}_{n_k}) +\cdots\right)+O(\frac{1}{c}), \nn\eea
where the summation over $m_j$ is from 2 to the infinity,  the summation over $n_k$ is from $3$ to the infinity, and $r_1,r-r_1$  are  the ``particle numbers'' for the inserting states coming from the vacuum module and $W_3$ module respectively. This completeness condition can be transformed into the one in terms of the vertex operators.

As in the pure gravity case, if we are interested in the two-point functions, we still have the completeness relation (\ref{complete2}) but now the single particle states should include the ones from the $W$ primaries.  Moreover, the states in different Verma modules are orthogonal to each other so that the two-point function of the vertex operators coming from different modules are vanishing. As a result, one may consider the contributions of different modules to the partition function (\ref{Zg}) separately and finally  multiply them together. Taking the $W_3$ module as an example, we find that the single-particle states in it contribute to the genus-$1$ partition function
\be
Z^{(1)}_{W}=\sum_{n=3}^\infty \langle ^{\cL}{\bar{W}}_n(r_1)W_n(a_1)\rangle =\Tr_{{\cal H}_{1,W}}q^{L_0}=\sum_{n=3}^\infty q^n
\ee
where ${\cal H}_{1,W}$ means the Hilbert space of the single-particle states in the $W_3$ module. For the multi-particle states, there are states like $\displaystyle{(\prod_{j=1}^{r_1}\hat{L}_{-m_j})(\prod_{k=1}^{r-r_1}\hat{W}_{-n_k})\mid 0\rangle}$. A multi-point function of these states on the Riemann sphere is factorized into the product of two-point functions, each of them being between the operators from the same module. With the completeness condition, only the operators from the same module can form link. Consequently, the final partition function is
\be
Z_g=Z_g^{(vacuum)}Z_g^{W_3}
\ee
where
\be
Z_g^{(vacuum)}=\prod_\g\left(\prod_{m=2}^\infty \frac{1}{1-q_\g^m}\right), \hs{3ex}Z_g^{W_3}=\prod_\g\left(\prod_{n=3}^\infty \frac{1}{1-q_\g^n}\right)
\ee
Therefore, the partition function $Z_g$ is exactly the same as the 1-loop partition function (\ref{1loops}) for the higher spin AdS$_3$ gravity.

\section{Conclusion and discussion}

In this paper, we discussed the 1-loop partition function in the AdS$_3$/CFT$_2$ correspondence. We focused on the handlebody solutions in the AdS$_3$ gravity.  These solutions end on the asymptotic boundary as compact Riemann surfaces, which could be described by Schottky uniformization. The 1-loop partition function (\ref{1loop}) of these solutions have been computed by using the heat kernel techniques and the method of images in \cite{Giombi:2008vd}. But the direct computation in the dual CFT has so far been missing. We filled this gap and proved the result (\ref{1loop}) in the large central charge limit of the CFT in this work.

We used the sewing technique to compute the CFT partition function on the  Riemann surface.  In the large $c$ limit, the leading contribution, which is linear in $c$ and corresponds to the semiclassical action of the gravitational configuration,  is captured by the ZT action. The sub-leading contribution to the partition function is encoded by the $2g$-point functions on the Riemann sphere. These multi-point functions are at most of order $c^0$ under the large $c$ limit. Actually it is relatively easy to read the leading order terms in these $2g$-point functions. It turns out that at leading order every $2g$-point function could be reduced to  the products of the two-point functions of single-particle operators. The products of two-point functions may define the links. Every oriented link is one-to-one related to a conjugacy class in the Schottky group. The  value of each link could be reduced to one two-point function, whose value is determined by the multiplier of the conjugacy element in the Schottky group. By considering all possible ways to contract the operators and form the links, the result (\ref{1loop}) has been reproduced. We generalized the study to the higher spin AdS$_3$ gravity and found agreement as well. 

The proof presented in this work relies on the essential fact that the dual CFT in the large $c$ limit is effectively free. As the two-point function of single-particle states dominates,  the contribution from three-point function is suppressed. As a result, the multi-point functions on the Riemann sphere is simplified. In the bulk side, the dominance of two-point function of the single-particle operator is reflected in the fact that the massless graviton is freely propagating and the interaction among gravitons can be ignored. 
Certainly, this should be the case since the 1-loop gravitational partition function is only given by the functional determinant of the free massless graviton. 

It would be interesting to study the higher loop partition function in the AdS$_3$ gravity from the multi-point functions on the Riemann sphere.  The recent study in \cite{Headrick:2015gba} shows that the higher order $1/c$ terms,  corresponding to the higher loop corrections, are not vanishing for higher genus  Riemann surface. 
It would be great to develop a systematic way to compute such terms in the large $c$ limit. However, this problem is rather difficult as several approximations we relied on have to be reconsidered carefully. First of all, the orthogonality of the different states in the vacuum module does not hold at the order $1/c$. Secondly, the vertex operators at the fixed points could not be written as the normal ordered product of single-particle operators. Moreover, besides two-point function, the three-point function of single-particle state should be taken into account. On the bulk side, this means that we have to consider the interaction of the gravitons. 

In this work, we mainly discussed the pure AdS$_3$ gravity and its higher spin generalization. It is easy to see that the 1-loop partition function in the chiral gravity\cite{Li:2008dq} can be proved. In this case, we only needs to consider the holomorphic sector of the CFT, then the result is implied in our discussion. For the topologically massive gravity\cite{Deser:1981wh} at critical point, there could be other consistent asymptotic boundary condition to allow the logarithmic modes so that the dual CFT is a logarithmic CFT\footnote{See \cite{Maloney:2009ck} for complete references on this issue.}. It would be interesting to check if we can reproduce the 1-loop partition function in this case and its higher spin generalization\cite{Gaberdiel:2010xv,Chen:2011vp,Bagchi:2011vr,Bagchi:2011td,Chen:2011yx}.


\vspace*{10mm}
\noindent {\large{\bf Acknowledgments}}\\

We would like to thank Jiang Long for valuable discussions, thank A. Maloney, I. Zadeh for usual conversations and correspondence. BC would like to thank the participants of the Workshop on ``Holography for black holes and cosmology" (ULB, 
Brussels), especially E. Perlmutter and T. Hartman, for  stimulating discussion, which initialized this project. The work was in part supported by NSFC Grant No.~11275010, No.~11335012 and No.~11325522. BC thanks Harvard University for hospitality during the final stage of this work. 
\vspace*{5mm}

\begin{appendix} 

\end{appendix}


\vspace*{5mm}
\begin{thebibliography}{99}

\bibitem{Maldacena:1997re} 
  J.~M.~Maldacena,
  ``The Large N limit of superconformal field theories and supergravity,''
  Int.\ J.\ Theor.\ Phys.\  {\bf 38}, 1113 (1999)
  [Adv.\ Theor.\ Math.\ Phys.\  {\bf 2}, 231 (1998)]
  [hep-th/9711200].

  \bibitem{Brown:1986nw}
  J.~D.~Brown and M.~Henneaux,
  ``Central Charges in the Canonical Realization of Asymptotic Symmetries: An Example from Three-Dimensional Gravity,''
  Commun.\ Math.\ Phys.\  {\bf 104}, 207 (1986).

  \bibitem{Banados:1992wn}
  M.~Banados, C.~Teitelboim and J.~Zanelli,
  ``The Black hole in three-dimensional space-time,''
  Phys.\ Rev.\ Lett.\  {\bf 69}, 1849 (1992)
  [hep-th/9204099].

  \bibitem{Banados:1992gq}
  M.~Banados, M.~Henneaux, C.~Teitelboim and J.~Zanelli,
  ``Geometry of the (2+1) black hole,''
  Phys.\ Rev.\ D {\bf 48}, 1506 (1993)
  [Phys.\ Rev.\ D {\bf 88}, no. 6, 069902 (2013)]
  [gr-qc/9302012].

  \bibitem{Strominger:1997eq}
  A.~Strominger,
  ``Black hole entropy from near horizon microstates,''
  JHEP {\bf 9802}, 009 (1998)
  [hep-th/9712251].

  \bibitem{Achucarro:1987vz}
  A.~Achucarro and P.~K.~Townsend,
  ``A Chern-Simons Action for Three-Dimensional anti-De Sitter Supergravity Theories,''
  Phys.\ Lett.\ B {\bf 180}, 89 (1986).

  \bibitem{Witten:1988hc}
  E.~Witten,
  ``(2+1)-Dimensional Gravity as an Exactly Soluble System,''
  Nucl.\ Phys.\ B {\bf 311}, 46 (1988).

  \bibitem{Witten:2007kt}
  E.~Witten,
  ``Three-Dimensional Gravity Revisited,''
  arXiv:0706.3359 [hep-th].

  \bibitem{Blencowe:1988gj}
  M.~P.~Blencowe,
  ``A Consistent Interacting Massless Higher Spin Field Theory in $D$ = (2+1),''
  Class.\ Quant.\ Grav.\  {\bf 6}, 443 (1989).

\bibitem{Bergshoeff:1989ns}
  E.~Bergshoeff, M.~P.~Blencowe and K.~S.~Stelle,
  ``Area Preserving Diffeomorphisms and Higher Spin Algebra,''
  Commun.\ Math.\ Phys.\  {\bf 128}, 213 (1990).

  \bibitem{Campoleoni:2010zq}
  A.~Campoleoni, S.~Fredenhagen, S.~Pfenninger and S.~Theisen,
  ``Asymptotic symmetries of three-dimensional gravity coupled to higher-spin fields,''
  JHEP {\bf 1011}, 007 (2010)
  [arXiv:1008.4744 [hep-th]].

  \bibitem{Henneaux:2010xg}
  M.~Henneaux and S.~J.~Rey,
  ``Nonlinear $W_{\infty}$ as Asymptotic Symmetry of Three-Dimensional Higher Spin Anti-de Sitter Gravity,''
  JHEP {\bf 1012}, 007 (2010)
  [arXiv:1008.4579 [hep-th]].

  \bibitem{Krasnov:2000zq}
  K.~Krasnov,
  ``Holography and Riemann surfaces,''
  Adv.\ Theor.\ Math.\ Phys.\  {\bf 4}, 929 (2000)
  [hep-th/0005106].

  \bibitem{Zagraf:1988}
P.G. Zograf and L. A. Takhtajan,
``On Uniformization of Riemann Surfaces and the Weil-Petersson Metric on Teichmuller and Schottky Spaces,''
   Math.\ USSR.\ Sb. {\bf 60} 297 (1988)

   \bibitem{Takhtajan:2002cc}
  L.~A.~Takhtajan and L.~P.~Teo,
  ``Liouville action and Weil-Petersson metric on deformation spaces, global Kleinian reciprocity and holography,''
  Commun.\ Math.\ Phys.\  {\bf 239}, 183 (2003)
  [math/0204318 [math-cv]].

  \bibitem{Yin:2007at}
  X.~Yin,
  ``On Non-handlebody Instantons in 3D Gravity,''
  JHEP {\bf 0809}, 120 (2008)
  [arXiv:0711.2803 [hep-th]].

  \bibitem{Yin:2007gv}
  X.~Yin,
  ``Partition Functions of Three-Dimensional Pure Gravity,''
  Commun.\ Num.\ Theor.\ Phys.\  {\bf 2}, 285 (2008)
  [arXiv:0710.2129 [hep-th]].

  \bibitem{Giombi:2008vd}
  S.~Giombi, A.~Maloney and X.~Yin,
  ``One-loop Partition Functions of 3D Gravity,''
  JHEP {\bf 0808}, 007 (2008)
  [arXiv:0804.1773 [hep-th]].

  \bibitem{Hartman:2013mia}
  T.~Hartman,
  ``Entanglement Entropy at Large Central Charge,''
  arXiv:1303.6955 [hep-th].

  \bibitem{Hartman:2014oaa}
  T.~Hartman, C.~A.~Keller and B.~Stoica,
  ``Universal Spectrum of 2d Conformal Field Theory in the Large c Limit,''
  JHEP {\bf 1409}, 118 (2014)
  [arXiv:1405.5137 [hep-th]].

  \bibitem{Maloney:2007ud}
  A.~Maloney and E.~Witten,
  ``Quantum Gravity Partition Functions in Three Dimensions,''
  JHEP {\bf 1002}, 029 (2010)
  [arXiv:0712.0155 [hep-th]].

  \bibitem{Faulkner:2013yia}
  T.~Faulkner,
  ``The Entanglement Renyi Entropies of Disjoint Intervals in AdS/CFT,''
  arXiv:1303.7221 [hep-th].

  \bibitem{Ryu:2006bv}
  S.~Ryu and T.~Takayanagi,
  ``Holographic derivation of entanglement entropy from AdS/CFT,''
  Phys.\ Rev.\ Lett.\  {\bf 96}, 181602 (2006)
  [hep-th/0603001].

  \bibitem{Ryu:2006ef}
  S.~Ryu and T.~Takayanagi,
  ``Aspects of Holographic Entanglement Entropy,''
  JHEP {\bf 0608}, 045 (2006)
  [hep-th/0605073].

  \bibitem{Headrick:2010zt}
  M.~Headrick,
  ``Entanglement Renyi entropies in holographic theories,''
  Phys.\ Rev.\ D {\bf 82}, 126010 (2010)
  [arXiv:1006.0047 [hep-th]].

  \bibitem{Barrella:2013wja}
  T.~Barrella, X.~Dong, S.~A.~Hartnoll and V.~L.~Martin,
  ``Holographic entanglement beyond classical gravity,''
  JHEP {\bf 1309}, 109 (2013)
  [arXiv:1306.4682 [hep-th]].

  \bibitem{Chen:2013kpa}
  B.~Chen and J.~J.~Zhang,
  ``On short interval expansion of R\'enyi entropy,''
  JHEP {\bf 1311}, 164 (2013)
  [arXiv:1309.5453 [hep-th]].

  \bibitem{Chen:2013dxa}
  B.~Chen, J.~Long and J.~j.~Zhang,
  ``Holographic R\'enyi entropy for CFT with W symmetry,''
  JHEP {\bf 1404}, 041 (2014)
  [arXiv:1312.5510 [hep-th]].

  \bibitem{Perlmutter:2013paa}
  E.~Perlmutter,
  ``Comments on Renyi entropy in AdS$_3$/CFT$_2$,''
  JHEP {\bf 1405}, 052 (2014)
  [arXiv:1312.5740 [hep-th]].

  \bibitem{Chen:2014kja}
  B.~Chen, F.~y.~Song and J.~j.~Zhang,
  ``Holographic Renyi entropy in AdS$_3$/LCFT$_2$ correspondence,''
  JHEP {\bf 1403}, 137 (2014)
  [arXiv:1401.0261 [hep-th]].

  \bibitem{Beccaria:2014lqa}
  M.~Beccaria and G.~Macorini,
  ``On the next-to-leading holographic entanglement entropy in $AdS_{3}/CFT_{2}$,''
  JHEP {\bf 1404}, 045 (2014)
  [arXiv:1402.0659 [hep-th]].

  \bibitem{Chen:2014unl}
  B.~Chen and J.~q.~Wu,
  ``Single interval Renyi entropy at low temperature,''
  JHEP {\bf 1408}, 032 (2014)
  [arXiv:1405.6254 [hep-th]].

  \bibitem{Chen:2015kua}
  B.~Chen and J.~q.~Wu,
  ``Holographic Calculation for Large Interval R\'enyi Entropy at High Temperature,''
  arXiv:1506.03206 [hep-th].

  \bibitem{Chen:2015uia}
  B.~Chen, J.~q.~Wu and Z.~c.~Zheng,
  ``Holographic R\'enyi Entropy of Single Interval on Torus: with W symmetry,''
  Phys.\ Rev.\ D {\bf 92}, 066002 (2015)
  [arXiv:1507.00183 [hep-th]].

  \bibitem{Segal:2002ei}
  G.~Segal,
  ``The definition of conformal field theory,'' in {\it Topology, geometry and quantum field theory},  {\bf 38} London Math. Soc. Lecture Note Ser., p.421ff. CUP, Cambridge (2004). 
                                                
                        
  \bibitem{Gaberdiel:2010jf}
  M.~R.~Gaberdiel, C.~A.~Keller and R.~Volpato,
  ``Genus Two Partition Functions of Chiral Conformal Field Theories,''
  Commun.\ Num.\ Theor.\ Phys.\  {\bf 4}, 295 (2010)
  [arXiv:1002.3371 [hep-th]].
  
  \bibitem{Headrick:2015gba}
  M.~Headrick, A.~Maloney, E.~Perlmutter and I.~G.~Zadeh,
  ``R\'enyi entropies, the analytic bootstrap, and 3D quantum gravity at higher genus,''
  JHEP {\bf 1507}, 059 (2015)
  [arXiv:1503.07111 [hep-th]].
  
  \bibitem{Fitzpatrick:2014vua} 
  A.~L.~Fitzpatrick, J.~Kaplan and M.~T.~Walters,
  ``Universality of Long-Distance AdS Physics from the CFT Bootstrap,''
  JHEP {\bf 1408}, 145 (2014)
  [arXiv:1403.6829 [hep-th]].
  
  \bibitem{Caputa:2014vaa} 
  P.~Caputa, M.~Nozaki and T.~Takayanagi,
  ``Entanglement of local operators in large-N conformal field theories,''
  PTEP {\bf 2014}, 093B06 (2014)
  [arXiv:1405.5946 [hep-th]].  
  
  

\bibitem{Li:2008dq} 
  W.~Li, W.~Song and A.~Strominger,
  ``Chiral Gravity in Three Dimensions,''
  JHEP {\bf 0804}, 082 (2008)
  [arXiv:0801.4566 [hep-th]].
 
 \bibitem{Deser:1981wh} 
  S.~Deser, R.~Jackiw and S.~Templeton,
  ``Topologically Massive Gauge Theories,''
  Annals Phys.\  {\bf 140}, 372 (1982)
  [Annals Phys.\  {\bf 185}, 406 (1988)]
  [Annals Phys.\  {\bf 281}, 409 (2000)]. 
  
\bibitem{Maloney:2009ck} 
  A.~Maloney, W.~Song and A.~Strominger,
  ``Chiral Gravity, Log Gravity and Extremal CFT,''
  Phys.\ Rev.\ D {\bf 81}, 064007 (2010)
  [arXiv:0903.4573 [hep-th]].

\bibitem{Gaberdiel:2010xv} 
  M.~R.~Gaberdiel, D.~Grumiller and D.~Vassilevich,
  ``Graviton 1-loop partition function for 3-dimensional massive gravity,''
  JHEP {\bf 1011}, 094 (2010)
  [arXiv:1007.5189 [hep-th]].
  
  \bibitem{Chen:2011vp} 
  B.~Chen, J.~Long and J.~B.~Wu,
  ``Spin-3 Topological Massive Gravity,''
  Phys.\ Lett.\ B {\bf 705}, 513 (2011)
  [arXiv:1106.5141 [hep-th]].
  
\bibitem{Bagchi:2011vr} 
  A.~Bagchi, S.~Lal, A.~Saha and B.~Sahoo,
  ``Topologically Massive Higher Spin Gravity,''
  JHEP {\bf 1110}, 150 (2011)
  [arXiv:1107.0915 [hep-th]].
  
 \bibitem{Bagchi:2011td} 
  A.~Bagchi, S.~Lal, A.~Saha and B.~Sahoo,
  ``One loop partition function for Topologically Massive Higher Spin Gravity,''
  JHEP {\bf 1112}, 068 (2011)
  [arXiv:1107.2063 [hep-th]].
        
\bibitem{Chen:2011yx} 
  B.~Chen and J.~Long,
  ``High Spin Topologically Massive Gravity,''
  JHEP {\bf 1112}, 114 (2011)
  [arXiv:1110.5113 [hep-th]].
  
\end{thebibliography}
\end{document}